\documentclass[12pt]{article}
\usepackage{a4wide}
\usepackage{latexsym}
\usepackage{amsmath}
\usepackage{amssymb}
\usepackage{amsfonts}
\usepackage{amscd}
\usepackage{amsthm}
\usepackage{cite}
\usepackage{placeins}

\usepackage{pslatex}
\usepackage{graphicx}
\usepackage[latin1,utf8]{inputenc}
\usepackage[T1]{fontenc}

\usepackage{mathtools}
\usepackage{mathdots}
\usepackage{bm}
\usepackage{nomencl}

\usepackage[dvipsnames]{xcolor}
\usepackage{tcolorbox}
\usepackage{colortbl}

\allowdisplaybreaks

\newcommand{\bq}{\begin{eqnarray}}
\newcommand{\eq}{\end{eqnarray}}
\newcommand{\eps}{\varepsilon}

\newcommand{\NB}{N}
\newcommand{\NF}{n}
\newcommand{\NL}{N_L}
\newcommand{\NS}{s}

\newcommand{\fieldK}{F}
\newcommand{\fieldL}{L}

\theoremstyle{plain}

\begin{document}

\thispagestyle{empty}

\begin{flushright}
  MITP/24-059 \\
  TUM-HEP-1513/24
\end{flushright}

\vspace{0.5cm}

\begin{center}
  {\Large\bf Self-dualities and Galois symmetries in Feynman integrals \\
  }
  \vspace{1cm}
  {\large Sebastian~P\"ogel${}^{a}$, Xing~Wang${}^{b}$, Stefan~Weinzierl${}^{a}$, Konglong~Wu${}^{c,d}$ and Xiaofeng~Xu${}^{a}$\\
  \vspace{1cm}
      {\small \em ${}^{a}$ PRISMA Cluster of Excellence,} 
      {\small \em Institut f{\"u}r Physik, Staudinger Weg 7,} \\
      {\small \em Johannes Gutenberg-Universit{\"a}t Mainz,}
      {\small \em D - 55099 Mainz, Germany}\\
  \vspace{2mm}
      {\small \em ${}^{b}$ Physik Department,} 
      {\small \em TUM School of Natural Sciences,} \\
      {\small \em Technische Universit\"at M\"unchen,} 
      {\small \em D - 85748 Garching, Germany} \\
  \vspace{2mm}
      {\small \em ${}^{c}$ Deutsches Elektronen-Synchrotron DESY,} 
      {\small \em Notkestrasse 85,} \\
      {\small \em D - 22607 Hamburg, Germany} \\
  \vspace{2mm}
      {\small \em ${}^{d}$ School of Physics and Technology,} 
      {\small \em Wuhan University, No.299 Bayi Road,} \\
      {\small \em Wuhan 430072, China}
  } 
\end{center}

\vspace{2cm}

\begin{abstract}\noindent
  {
It is well-known that all Feynman integrals within a given family can be expressed
as a finite linear combination of master integrals.
The master integrals naturally group into sectors.
Starting from two loops, there can exist sectors made up of more than one master integral.
In this paper we show that such sectors may have additional symmetries.
First of all, self-duality, which was first observed in Feynman integrals related to Calabi--Yau geometries,
often carries over to non-Calabi--Yau Feynman integrals.
Secondly, we show that in addition there can exist Galois symmetries relating integrals.
In the simplest case of two master integrals within a sector, whose definition involves a square root $r$,
we may choose a basis $(I_1,I_2)$ such that $I_2$ is obtained from $I_1$ by the substitution $r \rightarrow -r$. 
This pattern also persists in sectors, which a priori are not related to any square root with dependence on
the kinematic variables. 
We show in several examples that in such cases a suitable redefinition of the integrals introduces constant square roots like $\sqrt{3}$.
The new master integrals are then again related by a Galois symmetry, for example the substitution $\sqrt{3} \rightarrow -\sqrt{3}$.
To handle the case where the argument of a square root would be a perfect square
we introduce a limit Galois symmetry.
Both self-duality and Galois symmetries constrain the differential equation.
   }
\end{abstract}

\vspace*{\fill}

\newpage

\section{Introduction}

Integration-by-parts and differential equations are among the most popular tools 
to compute Feynman integrals.
Integration-by-parts allows us to express any integral from a family of Feynman integrals 
as a linear combination of master integrals \cite{Tkachov:1981wb,Chetyrkin:1981qh}. 
The set of master integrals is finite \cite{Smirnov:2010hn}.
The method of differential equations allows us to write down a differential equation for the master integrals
with respect to the kinematic variables \cite{Kotikov:1990kg,Kotikov:1991pm,Remiddi:1997ny,Gehrmann:1999as}.
In order to solve the differential equation one tries to find a transformation to an $\eps$-factorised form \cite{Henn:2013pwa}
\bq
 d I & = & \eps A I,
\eq
where $I$ denotes the vector of master integrals, the symbol $d$ is the differential with respect to all kinematic variables,
$\eps$ denotes the dimensional regularisation parameter, and $A$ is a square matrix with dimensions
equal to the number of master integrals. The entries of $A$ are differential one-forms,
depending on the kinematic variables, but independent of the dimensional regularisation parameter $\eps$. 
A differential equation in $\eps$-factorised form can be solved systematically order-by-order in $\eps$
in terms of iterated integrals \cite{Chen}.

In this paper we investigate the structure of the matrix $A$ in more detail.
In particular we show that it is often possible to redefine the master integrals, such that the $\eps$-factorised
form is maintained, but additional symmetries can be realised.
This is achieved by a transformation
\bq
 J & = & U I,
\eq
where $U$ is a $\mathrm{GL}(\NF,{\mathbb C})$-matrix, independent of the dimensional regularisation parameter $\eps$
and the kinematic variables. 
The number of master integrals is denoted by $\NF$.
Such transformations preserve the $\eps$-factorised form, as required.

Master integrals may be grouped into sectors. 
If we order the master integrals such that the first integral 
corresponds to the simplest and the last to the most complicated one, then the matrix $A$ has
a lower block-triangular structure.
The size of the blocks on the diagonal is given by the number of master integrals in the corresponding sector.
While at one loop, each sector has just one master integral, starting from two loops there may be sectors with more than one master integral.
In this paper we are interested in sectors with more than one master integral.
We show that it is often possible to redefine the integrals such that the master integrals within one sector 
exhibit a self-duality symmetry.
Self-duality is the statement that the block on the diagonal is reflection-symmetric with respect to the anti-diagonal.
This type of symmetry was first observed in Feynman integrals related to Calabi--Yau geometries \cite{Pogel:2022ken,Pogel:2022vat}.
Extending the analysis of ref.~\cite{Jiang:2023jmk}, we show that self-duality often extends to non-Calabi--Yau Feynman integrals.
In addition there can exist Galois symmetries, such that master integrals within the same sector are
related by the action of a Galois group.
In the simplest case such an action amounts to flipping the sign of a specific square root $r$ appearing in the integrals, i.e.~$r \to-r$.
Galois symmetries have been used in the past to group the letters of the symbol alphabet into even and odd letters, see for example refs.~\cite{Gehrmann:2015bfy,Chicherin:2017dob,FebresCordero:2023pww}.
In this paper we show that Galois symmetries often extend from symmetry properties of individual letters to relations between master integrals.

Both types of symmetries, self-duality and Galois symmetries, 
induce relations among the entries of the differential equation matrix $A$.
To give an example, consider a sector with two master integrals $I_1, I_2$, and assume that we may choose 
the two master integrals such that self-duality and Galois symmetry are manifest.
The block on the diagonal of the matrix $A$ (corresponding to the maximal cut of this sector)
\bq
 \left(\begin{array}{cc}
  a_{11} & a_{12} \\
  a_{21} & a_{22} \\
 \end{array} \right)
\eq
has then the self-duality symmetry
\bq
 a_{11} & = & a_{22}.
\eq
The Galois symmetry relates the two master integrals 
such that $I_2=\sigma(I_1)$,
where $\sigma$ is an element of the Galois group.
This induces on the matrix $A$ the relations
\bq
 a_{22} \; = \; \sigma\left(a_{11}\right),
 & &
 a_{21} \; = \; \sigma\left(a_{12}\right).
\eq
The combination of Galois symmetry and self-duality gives us therefore
\bq
 a_{22} \; = \; a_{11},
 \;\;\;\;\;\;
 a_{21} \; = \; \sigma\left(a_{12}\right),
 \;\;\;\;\;\;
 a_{11} \; = \; \sigma\left(a_{11}\right).
\eq
We see that we only need to know two entries of the block, the other two entries follow from symmetries.

Galois symmetries are expected if the definition of one master integral in the sector involves a square root $r$.
In the simplest case of two master integrals within one sector, whose definition involves a square root $r$,
we may choose a basis $(I_1,I_2)$ such that $I_2$ is obtained from $I_1$ by the substitution $r \rightarrow -r$. 
However, we can go further.
We point out that this pattern may also appear in sectors that a priori are not related to any square root with dependence on
the kinematic variables. 
We show that in these cases constant square roots such as $\sqrt{3}$ can be introduced, such that after a suitable redefinition the master integrals are related by the substitution $\sqrt{3} \rightarrow -\sqrt{3}$.

There is no reason to expect self-duality to extend to the full differential equation beyond a single sector:
If we order the master integrals by sectors and the sectors from the simplest to the most complicated, 
self-duality would relate the most complicated sector to the simplest one, which is obviously impossible.
However, we show in several examples that Galois symmetry does extend to the full matrix $A$. 
Consider a second sector consisting of two master integrals
$I_3, I_4$, and let us assume that they have been chosen such that $I_4=\sigma'(I_3)$, 
where $\sigma'$ is another element of the Galois group not identical to the first one. 
Let us further assume that $\sigma'$ acts trivially on $I_1$ and $I_2$, and that
$\sigma$ acts trivially on $I_3$ and $I_4$. 
The matrix $A$ is thus a $(4 \times 4)$-matrix.
The lower-left non-diagonal block
\bq
 \left(\begin{array}{cc}
  a_{31} & a_{32} \\
  a_{41} & a_{42} \\
 \end{array} \right)
\eq
has then the symmetries
\bq
 a_{32} \; = \; \sigma\left(a_{31}\right),
 \;\;\;\;\;\;
 a_{41} \; = \; \sigma'\left(a_{31}\right),
 \;\;\;\;\;\;
 a_{42} \; = \; \sigma'\left(\sigma\left(a_{31}\right)\right).
\eq
In this case we see that three entries out of four of the non-diagonal block may be obtained from Galois symmetries.
Our strategy is always to impose self-duality first, and in a second step to realise Galois symmetries.
Although we were able to make self-duality manifest in all examples we checked, we have no proof that this is possible in general.
While sectors with two or more master integrals often satisfy in addition to self-duality a Galois symmetry, 
we are aware of integrals where this is not the case. 
The most prominent examples are the massless planar double-box and the equal-mass sunrise integral.
In the case of the massless planar double-box integral we still have a ``limit Galois symmetry''.
We will explain this concept in detail in the main part of the paper.
In the case of the equal-mass sunrise integral we do not expect Galois symmetries, as Galois symmetries cannot 
relate quantities of different modular weight.

The transformation $U$, which realises self-duality and Galois symmetry is not necessarily unique.
We provide a simple example with two master integrals in one sector and two square roots $r_1$ and $r_2$
related to this sector.
We may either choose a basis such that $J_2=\sigma_1(J_1)$ (where $\sigma_1$ corresponds to $r_1 \rightarrow - r_1$)
or a basis $J_2'=\sigma_2(J_1')$ (where $\sigma_2$ corresponds to $r_2 \rightarrow - r_2$).
In a second example with three master integrals in one sector we show that there is even a 
one-parameter family of possible
transformations, which realise self-duality and Galois symmetry.

This paper is organised as follows:
In the following section we review the concept of master integrals, self-duality and the basics of Galois theory.
In section~\ref{sect:examples}
we show in several examples that it is possible to realise self-duality and Galois symmetries.
We first focus on examples, where the differential one-forms are dlog-forms with algebraic arguments.
We are in particular interested in the case, where square roots appear in the arguments of the dlog-forms.
For our study it is irrelevant, if the square roots can be rationalised simultaneously or not.
We present examples for both cases.
In section~\ref{sect:beyond_dlog} we comment on Feynman integrals, 
whose differential equations involve differential one-forms 
beyond dlog-forms with algebraic arguments.
This includes the elliptic case.
Finally, section~\ref{sect:conclusions} contains our conclusions.

\section{Set-up}
\label{sect:review}

\subsection{Master integrals}

We consider Feynman integrals, which depend on $\NB$ kinematic variables $x=(x_1,\dots,x_{\NB})$.
We view the kinematic variables as coordinates on the kinematic space $X$.
Let $I=(I_1,\dots,I_{\NF})^T$ be a vector of $\NF$ master integrals.
We assume that the master integrals satisfy an
$\eps$-factorised differential equation 
\bq
\label{diff_eq}
 d I\left(x,\eps\right) & = & \eps A\left(x\right) I\left(x,\eps\right).
\eq
with an integrable connection $A$:
\bq
 d A \; = \; 0
 & \mbox{and} &
 A \wedge A \; = \; 0,
\eq
In addition, we will always assume that when we restrict the kinematic variables to a sub-space, where the geometry
reduces to a curve of genus zero, the master integrals will be pure\footnote{This excludes for example the basis $K$ from ref.~\cite{Frellesvig:2023iwr}.}. 
The differential $d$ is the differential in the kinematic variables $x$:
\bq
 d & = & \sum\limits_{j=1}^{\NB} dx_j \frac{\partial}{\partial x_j}.
\eq
The condition $dA=0$ states that the entries of the $(\NF \times \NF)$-matrix $A$ are closed one-forms.
We denote a basis of differential one-forms appearing in $A$ by $\omega_1,\dots,\omega_{\NL}$
and the ${\mathbb C}$-vector space they span by $\Omega^1(X)$.
We have
\bq
 \dim \Omega^1(X) & = & \NL.
\eq
If we assume that the master integrals are ordered such that $I_1$ is the simplest and $I_{\NF}$ the most complicated, then the matrix $A$ has a lower block triangular structure induced by the sectors (or topologies) of the family
of Feynman integrals under consideration.
We distinguish blocks on the diagonal and blocks off the diagonal.
A block on the diagonal corresponds to the maximal cut of the corresponding sector.
If a family of Feynman integrals has $\NS$ sectors then the matrix $A$ is of the form
\bq
 A 
 & = &
 \left( \begin{array}{ccccc}
  D_1 & 0 & \ldots & \ldots & 0 \\
  N_{21} & D_2 & \ddots & & \vdots \\
  \vdots & \ddots & \ddots & \ddots & \vdots \\
  \vdots & & \ddots & D_{\NS-1} & 0 \\
  N_{\NS 1} & \ldots & \ldots & N_{\NS (\NS-1)} & D_{\NS} \\
 \end{array}
 \right),
\eq
with diagonal blocks $D_i$ and non-diagonal blocks $N_{ij}$.

At one loop, every sector has just one master integral, but starting from two loops we may have sectors with two or more master integrals.
In this paper we are primarily interested in sectors of the second type.

\subsection{Self-duality}

Let us consider an $(n\times n)$ diagonal block
\bq
\label{self_duality_diagonal_block}
 D
 & = &
 \left( \begin{array}{ccccc}
  \cellcolor{red} d_{11} & \cellcolor{orange} d_{12} & \dots & \cellcolor{green} d_{1(n-1)} & d_{1n} \\
  \cellcolor{yellow} d_{21} & \cellcolor{cyan} d_{22} & \dots & d_{2(n-1)} & \cellcolor{green} d_{2n} \\
  \vdots & \vdots & & \vdots & \vdots \\
  \cellcolor{lime} d_{(n-1)1} & d_{(n-1)2} & \dots & \cellcolor{cyan} d_{(n-1)(n-1)} & \cellcolor{orange} d_{(n-1)n} \\
  d_{n1} & \cellcolor{lime} d_{n2} & \dots & \cellcolor{yellow} d_{n(n-1)} & \cellcolor{red} d_{nn} \\
 \end{array} \right).
\eq
Self-duality is the statement
\bq
\label{self_duality}
 d_{ij} & = & d_{(n+1-j)(n+1-i)}.
\eq
It corresponds to a reflection symmetry with respect to the anti-diagonal, as indicated by the colours in eq.~(\ref{self_duality_diagonal_block}).

Self-duality has been observed for the first time in the $l$-loop banana integrals of equal mass \cite{Pogel:2022yat,Pogel:2022ken,Pogel:2022vat}.
These integrals are related to Calabi--Yau geometries and the name derives from self-dual properties of Calabi--Yau operators \cite{2013arXiv1304.5434B}.
However, the symmetry stated in eq.~(\ref{self_duality}) is more general and not necessarily tied to Calabi--Yau geometries \cite{Jiang:2023jmk}.

Self-duality is a symmetry of the diagonal blocks.
There is no reason to expect self-duality to hold beyond the diagonal blocks if we keep the ordering:
If we order the master integrals by sectors, and the sectors from the simplest to the most complicated, 
self-duality would relate the most complicated sector to the simplest one, which is obviously impossible.

\subsection{Galois theory}

Given a non-constant polynomial $p(x)$ with coefficients from a field $\fieldK$, the roots of $p(x)$ may not lie in $\fieldK$.
In this case one considers the splitting field $\fieldL/\fieldK$, which is the smallest field extension that contains all the roots of $p(x)$.
The Galois group 
\bq
 G\left( \fieldL / \fieldK \right)
 & = & 
 \left\{ \; \sigma \in \mathrm{Aut}\left( \fieldL \right) \; | \; \left. \sigma \right|_{\fieldK} = \mathrm{id} \; \right\}
\eq
is the subgroup of the automorphism group of $\fieldL$, which keeps $\fieldK$ fixed.

A trivial example is given by the polynomial $p(x)=x^2-3 \in {\mathbb Q}[x]$.
The roots of $p(x)$ lie in ${\mathbb Q}[\sqrt{3}]$
and the Galois group is
\bq
 G\left( {\mathbb Q}[\sqrt{3}] / {\mathbb Q} \right)
 & = &
 {\mathbb Z}_2,
\eq
generated by
\bq
 \sigma & : & {\mathbb Q}[\sqrt{3}] \rightarrow {\mathbb Q}[\sqrt{3}],
 \nonumber \\
 & & \sigma\left(\sqrt{3}\right) \; = \; - \sqrt{3}.
\eq
In the application towards Feynman integrals we often encounter roots $r$ of quadratic equations, where 
the Galois group acts as $r \rightarrow -r$.
A typical example is the square root
\bq
 r & = & \sqrt{-s\left(4m^2-s\right)}.
\eq
In the differential equation we will have differential one-forms that are even, like
\bq
 \omega_0 & = & d\ln\left(\frac{s}{\mu^2}\right)
\eq
and differential one-forms that are odd, like
\bq
 \omega_1 & = & \frac{1}{2} d\ln\left(\frac{2m^2-s-r}{2m^2-s+r}\right).
\eq
The group element $\sigma$ sending $r$ to $(-r)$ acts on these as
\bq
 \sigma\left(\omega_0\right) \; = \; \omega_0,
 & &
 \sigma\left(\omega_1\right) \; = \; -\omega_1.
\eq
If two master integrals $J_1$ and $J_2$ are related by $J_2=\sigma(J_1)$ with $\sigma^2=\mathrm{id}$ it follows 
from
\bq
 d J_1 \; = \; \eps \sum\limits_{j=1}^{\NF} a_{1j} J_j,
 & &
 d J_2 \; = \; \eps \sum\limits_{j=1}^{\NF} a_{2j} J_j
\eq
that 
\bq
 \sigma\left(a_{12}\right)
 \; = \; a_{21},
 & &
 \sigma\left(a_{11}\right)
 \; = \; a_{22}.
\eq
If in addition the remaining master integrals are invariant under the action of $\sigma$, i.e. $\sigma(J_k)=J_k$
for $k \in \{3,\dots,\NF\}$ we further have
\bq
 \sigma\left(a_{1k}\right)
 \; = \;
 a_{2k}
 & \mbox{for} &
 k \; \in \; \left\{3,\dots,\NF\right\}.
\eq
More formally, we may view a system of Feynman integrals as a vector bundle. 
The base space is parameterised by the kinematic variables $x$, and for each point in the base space
we have a vector space in the fibre.
This vector space is spanned by the master integrals.
Initially we may take this vector space to be defined over the field ${\mathbb Q}(x,\eps)$, the field of
rational functions with rational coefficients in the kinematic variables $x$ and the dimensional regularisation
parameter $\eps$.
In a pre-canonical basis this is all what is needed: In the differential equation for this basis we will only have
rational functions in $x$ and $\eps$.
However, we are interested in an $\eps$-factorised basis and this may require to enlarge the field, for
example by adjoining a root $r$. In this case we are led to a vector space over the field ${\mathbb Q}(x,\eps)[r]$.
Throughout this paper we will assume that any element of the Galois group acts 
trivially on any pre-canonical master integral:
\bq
 \sigma\left(K\right) & = & K,
\eq
where $\sigma$ is an element of the Galois group and $K$ a pre-canonical master integral.

\subsection{Combination of Galois symmetries and self-duality}

Given a sector with two master integrals it is always possible to impose a Galois symmetry (without requiring in addition self-duality):
For two master integrals $I=(I_1,I_2)^T$ satisfying the $\eps$-factorised differential equation
\bq
 d I & = & \eps \tilde{A} I,
 \;\;\;\;\;\;\;\;\;
 \tilde{A} \; = \;
 \left( \begin{array}{cc}
 \tilde{a}_{11} & \tilde{a}_{12} \\
 \tilde{a}_{21} & \tilde{a}_{22} \\
 \end{array} \right),
\eq
one sets
\bq
 J_1 & = & I_1 + r I_2,
 \nonumber \\
 J_2 & = & I_1 - r I_2,
\eq
where $r$ is an algebraic extension of ${\mathbb Q}$, for example $r=i$.
We define the Galois action by $\sigma(r)=-r$, and $\sigma$ acts trivially on all other expressions.
Clearly we have
\bq
 J_2 & = & \sigma\left(J_1\right).
\eq
In the new basis $J=(J_1,J_2)^T$ the differential equation reads
\bq
 d J & = & \eps A J
\eq
with
\bq
 A
 \; = \;
 \left( \begin{array}{cc}
 a_{11} & a_{12} \\
 a_{21} & a_{22} \\
 \end{array} \right)
 \; = \; 
 \frac{1}{2}
 \left( \begin{array}{cc}
 \tilde{a}_{11}+\tilde{a}_{22} + r^{-1} \tilde{a}_{12} + r \tilde{a}_{21} & \tilde{a}_{11}-\tilde{a}_{22} - r^{-1} \tilde{a}_{12} + r \tilde{a}_{21} \\
 \tilde{a}_{11}-\tilde{a}_{22} + r^{-1} \tilde{a}_{12} - r \tilde{a}_{21} & \tilde{a}_{11}+\tilde{a}_{22} - r^{-1} \tilde{a}_{12} - r \tilde{a}_{21} \\
 \end{array} \right).
\eq
The matrix $A$ has the Galois symmetries
\bq
 a_{22} \; = \; \sigma\left(a_{11}\right),
 & &
 a_{21} \; = \; \sigma\left(a_{12}\right).
\eq
However, $A$ is in general not self-dual since
\bq
 a_{11} \; \neq \; \sigma\left(a_{11}\right).
\eq
Self-duality of $A$ will require
\bq
 \tilde{a}_{12} + r^2 \tilde{a}_{21} & = & 0.
\eq
The possibility of finding a basis which makes self-duality and Galois symmetries manifest is therefore a non-trivial property.
As Galois symmetry alone is trivial, we are always interested in the case where we might have Galois symmetries in addition to self-duality.

Let us summarise: For a sector with two master integrals it is often possible to find 
a basis $J=(J_1, J_2)^T$ such that 
\bq
 d J & = & \eps A J,
\eq
and $A$ has the structure
\bq
 A
 & = &
 \left( \begin{array}{cc}
 \cellcolor{red} a_{11} & \cellcolor{orange} a_{12} \\
 \cellcolor{orange} a_{21} & \cellcolor{red} a_{22} \\
 \end{array} \right),
\eq
where entries with the same background colour are related by a symmetry.
Self-duality relates
\bq
\label{symmetry_1}
 a_{11} & = & a_{22}.
\eq
Furthermore, there is a ${\mathbb Z}_2$-group with a generator $\sigma$ such that
\bq
\label{symmetry_2}
 a_{11} \; = \; \sigma\left(a_{11}\right),
 \;\;\;\;\;\;
 a_{12} \; = \; \sigma\left(a_{21}\right),
 \;\;\;\;\;\;
 a_{21} \; = \; \sigma\left(a_{12}\right),
 \;\;\;\;\;\;
 a_{22} \; = \; \sigma\left(a_{22}\right).
\eq
This is the Galois symmetry.
We see that only two of the four entries of the $(2 \times 2)$-matrix need to be known, the other two 
follow from symmetries.

It is worth discussing these concepts for sectors with more than two master integrals.
Let us consider a sector with three master integrals $I=(I_1,I_2,I_3)^T$ 
and the $\eps$-factorised differential equation
\bq
 d I & = & \eps A I,
 \;\;\;\;\;\;\;\;\;
 A \; = \;
 \left( \begin{array}{ccc}
 a_{11} & a_{12} & a_{13} \\
 a_{21} & a_{22} & a_{23} \\
 a_{31} & a_{32} & a_{33} \\
 \end{array} \right).
\eq
Self-duality gives the three relations
\bq
 a_{33} \; = \; a_{11},
 \;\;\;\;\;\;
 a_{23} \; = \; a_{12},
 \;\;\;\;\;\;
 a_{32} \; = \; a_{21}.
\eq
Let us assume that there exists a Galois symmetry, which relates $I_1$ and $I_3$
\bq
 I_3 & = & \sigma\left(I_1\right),
\eq
and which acts trivially on $I_2$:
\bq
\label{Galois_action_middle_master_integral}
 I_2 & = & \sigma\left(I_2\right).
\eq
The Galois symmetry alone leads to the relations
\bq
 a_{31} \; = \; \sigma\left(a_{13}\right),
 \;\;\;\;\;\;
 a_{32} \; = \; \sigma\left(a_{12}\right),
 \;\;\;\;\;\;
 a_{33} \; = \; \sigma\left(a_{11}\right),
 \;\;\;\;\;\;
 a_{21} \; = \; \sigma\left(a_{23}\right),
\eq
and to the invariance relation $a_{22} = \sigma(a_{22})$.
Combining self-duality and Galois symmetry we obtain the relations
\bq
 a_{33} \; = \; a_{11},
 \;\;\;\;\;\;
 a_{23} \; = \; a_{12},
 \;\;\;\;\;\;
 a_{32} \; = \; a_{21} \; = \; \sigma\left(a_{12}\right),
 \;\;\;\;\;\;
 a_{31} \; = \; \sigma\left(a_{13}\right),
\eq
and
\bq
 a_{11} \; = \; \sigma\left(a_{11}\right),
 \;\;\;\;\;\;
 a_{22} \; = \; \sigma\left(a_{22}\right).
\eq
This is illustrated in fig.~\ref{fig_symmetries}.
\begin{figure}
\begin{center}
\bq
 \left( \begin{array}{ccc}
 \cellcolor{red} a_{11} & \cellcolor{orange} a_{12} & \cellcolor{yellow} a_{13} \\
 \cellcolor{green} a_{21} & \cellcolor{cyan} a_{22} & \cellcolor{orange} a_{23} \\
 \cellcolor{lime} a_{31} & \cellcolor{green} a_{32} & \cellcolor{red} a_{33} \\
 \end{array} \right) 
\;\;\;\;\;\;\;\;\;
 \left( \begin{array}{ccc}
 \cellcolor{red} a_{11} & \cellcolor{orange} a_{12} & \cellcolor{yellow} a_{13} \\
 \cellcolor{green} a_{21} & \cellcolor{cyan} a_{22} & \cellcolor{green} a_{23} \\
 \cellcolor{yellow} a_{31} & \cellcolor{orange} a_{32} & \cellcolor{red} a_{33} \\
 \end{array} \right) 
\;\;\;\;\;\;\;\;\;
 \left( \begin{array}{ccc}
 \cellcolor{red} a_{11} & \cellcolor{green} a_{12} & \cellcolor{yellow} a_{13} \\
 \cellcolor{green} a_{21} & \cellcolor{cyan} a_{22} & \cellcolor{green} a_{23} \\
 \cellcolor{yellow} a_{31} & \cellcolor{green} a_{32} & \cellcolor{red} a_{33} \\
 \end{array} \right) 
 \nonumber
\eq
\end{center}
\caption{
The effect of various symmetries on a $(3 \times 3)$ diagonal block: 
Entries with the same background colour are related by a symmetry.
Left: Self-duality symmetry.
Middle: The Galois symmetry $I_3=\sigma(I_1)$, $I_2=\sigma(I_2)$.
Right: The combination of both.
}
\label{fig_symmetries}
\end{figure}
Combining self-duality and Galois symmetry, only four out of the nine entries of the matrix $A$ need to be known,
the remaining ones follow from symmetry.

The Galois symmetries extend beyond the maximal cut.
To discuss this point let us consider a system consisting of two sectors with two master integrals each, satisfying an $\eps$-factorised differential equation
\bq
 d I & = & \eps A I,
 \;\;\;\;\;\;
 A \; = \;
 \left( \begin{array}{cccc}
 a_{11} & a_{12} & 0 & 0 \\
 a_{21} & a_{22} & 0 & 0 \\
 a_{31} & a_{32} & a_{33} & a_{34} \\
 a_{41} & a_{42} & a_{43} & a_{44} \\
 \end{array} \right).
\eq
If the master integrals have been chosen such that self-duality is manifest, we have
\bq
 a_{22} \; = \; a_{11}
 & \mbox{and} &
 a_{44} \; = \; a_{33}.
\eq
Let us further assume that there are Galois group elements $\sigma$ and $\sigma'$, which relate $I_1, I_2$ and $I_3, I_4$, respectively.
In other words,
\bq
 I_2 \; = \; \sigma\left(I_1\right)
 & \mbox{and} &
 I_4 \; = \; \sigma'\left(I_3\right).
\eq
We then have
\bq
 a_{21} \; = \; \sigma\left(a_{12} \right),
 & &
 a_{43} \; = \; \sigma'\left(a_{34} \right).
\eq
Two cases are relevant: 
Within the first case, $\sigma$ acts trivially on $I_3, I_4$
\bq
 \sigma\left(I_3\right) \; = \; I_3
 & \mbox{and} &
 \sigma\left(I_4\right) \; = \; I_4,
\eq
and $\sigma'$ acts trivially on $I_1, I_2$
\bq
 \sigma'\left(I_1\right) \; = \; I_1
 & \mbox{and} &
 \sigma'\left(I_2\right) \; = \; I_2.
\eq
In this case the entries of the lower-left non-diagonal block are related
as
\bq
\label{symmetry_3}
 a_{32} \; = \; \sigma\left(a_{31}\right),
 \;\;\;\;\;\;\;\;\;
 a_{41} \; = \; \sigma'\left(a_{31}\right),
 \;\;\;\;\;\;\;\;\;
 a_{42} \; = \; \sigma'\left(\sigma\left(a_{31}\right)\right).
\eq
We see that in this case we only need to know one entry of the non-diagonal block, 
the other three follow from symmetry.
In summary, we have in this case the following structure of the matrix $A$
\bq
 A
 & = &
 \left( \begin{array}{cccc}
 \cellcolor{red} a_{11} & \cellcolor{orange} a_{12} & 0 & 0 \\
 \cellcolor{orange} a_{21} & \cellcolor{red} a_{22} & 0 & 0 \\
 \cellcolor{yellow} a_{31} & \cellcolor{yellow} a_{32} & \cellcolor{green} a_{33} & \cellcolor{lime} a_{34} \\
 \cellcolor{yellow} a_{41} & \cellcolor{yellow} a_{42} & \cellcolor{lime} a_{43} & \cellcolor{green} a_{44} \\
 \end{array} \right).
\eq
where entries with the same background colour are related by a symmetry.

The second relevant case is $\sigma=\sigma'$, i.e. the group element relates $I_1$ to $I_2$ as well as
$I_3$ to $I_4$.
We look again at the lower-left non-diagonal block.
In this case we have
\bq
\label{symmetry_4}
 a_{42} \; = \; \sigma\left(a_{31}\right),
 & &
 a_{41} \; = \; \sigma\left(a_{32}\right),
\eq
and two entries of the non-diagonal block may be obtained from Galois symmetries.
The structure of the matrix $A$ is then
\bq
 A
 & = &
 \left( \begin{array}{cccc}
 \cellcolor{red} a_{11} & \cellcolor{orange} a_{12} & 0 & 0 \\
 \cellcolor{orange} a_{21} & \cellcolor{red} a_{22} & 0 & 0 \\
 \cellcolor{magenta} a_{31} & \cellcolor{pink} a_{32} & \cellcolor{green} a_{33} & \cellcolor{lime} a_{34} \\
 \cellcolor{pink} a_{41} & \cellcolor{magenta} a_{42} & \cellcolor{lime} a_{43} & \cellcolor{green} a_{44} \\
 \end{array} \right),
\eq
where again entries with the same background colour are related by a symmetry.

The non-diagonal blocks need not be square matrices.
In the case where we have one sector with one master integral and a second sector with two master integrals
together with a generator $\sigma$ of the Galois group, the structure of the matrix $A$ in a suitable basis 
is given by
\bq
 A
 & = &
 \left( \begin{array}{ccc}
 a_{11} & 0 & 0 \\
 \cellcolor{yellow} a_{21} & \cellcolor{green} a_{22} & \cellcolor{lime} a_{23} \\
 \cellcolor{yellow} a_{31} & \cellcolor{lime} a_{32} & \cellcolor{green} a_{33} \\
 \end{array} \right).
\eq
Self-duality relates $a_{33} = a_{22}$, the Galois symmetry relates on the diagonal block $a_{32}=\sigma(a_{23})$.
On the $(2 \times 1)$-non-diagonal block the Galois symmetry relates
\bq
 a_{31} & = & \sigma\left(a_{21}\right).
\eq
The situation is similar if we have one sector with two master integrals and one sector with one master integral
(ordered from the simplest to the most complicated sector).
The structure of the matrix $A$ in a suitable basis is
\bq
 A
 & = &
 \left( \begin{array}{ccc}
 \cellcolor{red} a_{11} & \cellcolor{orange} a_{12} & 0 \\
 \cellcolor{orange} a_{21} & \cellcolor{red} a_{22} & 0 \\
 \cellcolor{yellow} a_{31} & \cellcolor{yellow} a_{32} & a_{33} \\
 \end{array} \right).
\eq
Self-duality relates $a_{22} = a_{11}$, the Galois symmetry relates on the diagonal block $a_{21}=\sigma(a_{12})$.
On the $(1 \times 2)$-non-diagonal block the Galois symmetry relates
\bq
 a_{32} & = & \sigma\left(a_{31}\right).
\eq

\subsection{Galois symmetries and rationalisations}
\label{sect:rationalisation}

Certain square roots can be rationalised \cite{Besier:2018jen,Besier:2019kco}
and one may consider the fate of Galois symmetries if 
one does so.
We discuss this case with a simple example, involving the square root
\bq
 r & = & \sqrt{-v\left(4-v\right)}.
\eq
Typical differential one-forms in this case are
\bq
 \omega_1
 \; = \; 
 d\ln\left(-v\right)
 \;\;\;\;\;\;
 \omega_2
 \; = \;
 d\ln\left(4-v\right),
 \;\;\;\;\;\;
 \omega_3
 \; = \;
 \frac{1}{2} d \ln \frac{2-v-r}{2-v+r}.
\eq
Let $\sigma$ be the element of the Galois group, which sends $r \rightarrow -r$.
The first two differential one-forms are even under $\sigma$, the third one is odd:
\bq
 \sigma\left(\omega_1\right) \; = \; \omega_1,
 \;\;\;\;\;\;
 \sigma\left(\omega_2\right) \; = \; \omega_2,
 \;\;\;\;\;\;
 \sigma\left(\omega_3\right) \; = \; -\omega_3.
\eq
The root $r$ is rationalised by
\bq
 v \; = \; - \frac{\left(1-x\right)^2}{x},
 \;\;\;\;\;\;
 x \; = \; \frac{1}{2} \left( 2 - v - r\right),
 \;\;\;\;\;\;
 r \; = \; \frac{1-x^2}{x}.
\eq
In the variable $x$ we have
\bq
 \omega_1
 \; = \; 
 2 d\ln\left(1-x\right) - d\ln x,
 \;\;\;\;\;\;
 \omega_2
 \; = \;
 2 d\ln\left(1+x\right) - d\ln x,
 \;\;\;\;\;\;
 \omega_3
 \; = \;
 d\ln x.
\eq
The transformation $\sigma(r)=-r$ translates to $\sigma(x)=x^{-1}$.
It is easily checked that $\omega_1$ and $\omega_2$ are invariant under $x \rightarrow x^{-1}$, while $\omega_3$
changes the sign under this transformation\footnote{The condition $\sigma(r)=-r$ leads in $x$-space to
the two possible transformations $x \rightarrow x^{-1}$ and $x \rightarrow -x$. However, only the first one leaves $\omega_1$ and $\omega_2$ invariant.}.
Thus we see that the automorphism $\sigma$ of ${\mathbb Q}(v,\eps)[r]$, which keeps ${\mathbb Q}(v,\eps)$ fixed
and sends $r \rightarrow -r$ corresponds to the automorphism $x \rightarrow x^{-1}$ of ${\mathbb Q}(x,\eps)$.
Note that in the latter case there is no Galois extension.

\subsection{Parameterised Galois symmetries and limit Galois symmetries}
\label{sect:limit_Galois_symmetry}

In some cases the transformation that realises self-duality and Galois symmetries is not unique.
For example, it might occur that the transformation
\bq
 J_1 & = & I_1 + r I_2,
 \;\;\;\;\;\;
 r \; = \; \sqrt{\lambda},
 \nonumber \\
 J_2 & = & I_1 - r I_2,
\eq
realises self-duality and Galois symmetry for any value $\lambda \in {\mathbb Q}$ that is not a perfect square.
We call this a parameterised Galois symmetry.
An example will be given in section~\ref{sect:pentabox}.
We have to exclude the case where $\lambda$ is a perfect square.
If $\lambda$ is a perfect square, we have $r \in {\mathbb Q}$ and there is no field extension.
Furthermore, there exists for $r \in \mathbb Q$ no field automorphism of ${\mathbb Q}$ which sends $r$ to $-r$.
 
Nevertheless, it will be useful to introduce something which comes close to being a Galois symmetry with respect
to a square root of a perfect square, as there will be cases where there is no (normal) Galois symmetry on top of self-duality.
We introduce the concept of a limit Galois symmetry as follows:
We first divide the rational numbers into
\bq
 {\mathbb Q}
 & = &
 {\mathbb P}{\mathbb S}
 \; \cup \;
 {\mathbb N}{\mathbb P}{\mathbb S},
\eq
where ${\mathbb P}{\mathbb S}$ denotes the set of rational numbers that are
perfects squares and ${\mathbb N}{\mathbb P}{\mathbb S}$
the set of rational numbers that are not perfect squares.
We consider sequences
\bq
 \left( \lambda_n \right) & \in & {\mathbb N}{\mathbb P}{\mathbb S}
\eq
with
\bq
 \lim\limits_{n \rightarrow \infty} \lambda_n & = & \lambda \; \in \; {\mathbb P}{\mathbb S}.
\eq
For each such sequence and each $\lambda_n$ we consider the field extension ${\mathbb Q}(x,\eps)[\sqrt{\lambda_n}]$
and redefine the master integrals for example as
\bq
 J^{(n)}_1 \; = \; I_1 + \sqrt{\lambda_n} I_2,
 & &
 J^{(n)}_2 \; = \; I_1 - \sqrt{\lambda_n} I_2.
\eq
For any $\lambda_n \in {\mathbb N}{\mathbb P}{\mathbb S}$ the master integrals 
$J^{(n)}_1$ and $J^{(n)}_2$ are related by a Galois symmetry
\bq
 J^{(n)}_2 & = & \sigma\left(J^{(n)}_1\right),
\eq
where $\sigma$ sends $\sqrt{\lambda_n}$ to $-\sqrt{\lambda_n}$.
We further set
\bq
 J_1 & = & \lim\limits_{n \rightarrow \infty}  J^{(n)}_1 \; = \; I_1 + \sqrt{\lambda} I_2,
 \nonumber \\
 J_2 & = & \lim\limits_{n \rightarrow \infty}  J^{(n)}_2 \; = \; I_1 - \sqrt{\lambda} I_2.
\eq
We say that $J_1$ and $J_2$ are related by a limit Galois symmetry, i.e. for any sequence $(\lambda_n)$
of non-perfect squares which converges to the perfect square $\lambda$ we have for any $\lambda_n$ a Galois
symmetry in the usual sense.

Our most important example will be a sequence $(\lambda_n)$ which converges to $1$.
We may take
\bq
 \lambda_n & = & 1 - \frac{1}{p_n},
\eq
where $p_n$ denotes the $n$-th prime number.
Clearly, $\lambda_n$ is not a perfect square and $\lim\limits_{n\rightarrow \infty} \lambda_n = 1$.
It will be convenient to use a short-hand notation for a limit Galois symmetry, mirroring the one we use
for a normal Galois symmetry.
In the following we will for example simply write
\bq
 J_1 \; = \; I_1 + r I_2,
 & &
 J_2 \; = \; I_1 - r I_2,
 \;\;\;\;\;\;
 r \; = \; \sqrt{1},
\eq
where $\sqrt{1}$ is understood in the sense discussed above.
In the combination of self-duality and a limit Galois symmetry we will require
self-duality only in the limit $n \rightarrow \infty$.
In section~\ref{sect:example_doublebox} we will present an example 
where there is no (normal) Galois symmetry in addition to self-duality.
However, there is a limit Galois symmetry in addition to self-duality.

\section{Examples with dlog-forms}
\label{sect:examples}

In this section we show in several examples that it is often possible to realise self-duality and Galois symmetries.
For now we limit the discussion to Feynman integrals whose $\eps$-factorised differential equation involves only dlog-forms with algebraic arguments.
Typically we will have square roots appearing in the arguments of the dlog-forms.
For the study of symmetries it is irrelevant whether the square roots can be rationalised simultaneously or not.

Examples of Feynman integrals with $\eps$-factorised differential equations beyond dlog-forms will be discussed in section~\ref{sect:beyond_dlog}.

\subsection{Drell--Yan}
\label{sect:drell_yan}

\subsubsection{Set-up}

As our main example, which we discuss at length, serves 
a diagram contributing to the mixed QCD-electroweak corrections to the Drell--Yan process.
This example involves three square roots.
It is known that the three square roots cannot be rationalised simultaneously \cite{Besier:2019hqd}.
However, the result can be expressed in terms of multiple polylogarithms \cite{Heller:2019gkq}.
The example we discuss is the minimal example with three non-simultaneously rationalisable square roots.
This section also serves to set-up our notation for all further examples.

We consider the integrals
\bq
 I_{\nu_1 \nu_2 \nu_3 \nu_4 \nu_5 \nu_6 \nu_7 \nu_8 \nu_9}
 = e^{2 \gamma_E \eps}
 \left(\mu^2\right)^{\nu-D}
 \int \left( \prod\limits_{a=1}^{2} \frac{d^Dk_a}{i \pi^{\frac{D}{2}}} \right)
 \left( \prod\limits_{c=1}^{9} \frac{1}{P_c^{\nu_c}} \right),
\eq
where $D$ denotes the number of space-time dimensions,
$\eps$ the dimensional regularisation parameter,
$\gamma_E$ the Euler--Mascheroni constant
and 
\bq
 \nu & = & \sum\limits_{j=1}^{9} \nu_j.
\eq 
The inverse propagators are given by
\begin{align}
 P_1 & = -k_2^2,
 &
 P_2 & = -k_1^2 + m^2,
 &
 P_3 & = -\left(k_1-k_2\right)^2,
 \nonumber \\
 P_4 & = -\left(k_2+p_1\right)^2,
 &
 P_5 & = -\left(k_1+p_1\right)^2,
 &
 P_6 & = -\left(k_2+p_1+p_2\right)^2,
 \nonumber \\
 P_7 & = -\left(k_1+p_1+p_2\right)^2 + m^2,
 & 
 P_8 & = -\left(k_2-p_4\right)^2,
 &
 P_9 & = -\left(k_1-p_4\right)^2.
\end{align}
The external particles are assumed to be massless: $p_1^2=p_2^2=p_3^2=p_4^2=0$.
The Mandelstam variables $s$ and $t$ are defined by
\bq
\label{def_Mandelstam}
 s \; = \; \left(p_1+p_2\right)^2,
 & &
 t \; = \; \left(p_2+p_3\right)^2.
\eq
A sector is defined by the set of propagators with positive exponents.
We define the sector id by
\bq
\label{def_sector_id}
 \mathrm{ID}
 & = & \sum\limits_{j=1}^9 2^{j-1} \Theta(\nu_j),
\eq
with $\Theta(x)=1$ for $x>0$ and $\Theta(x)=0$ for $x\le0$.
We further define the dimension-shift operator ${\bf D}^-$,
which lowers the dimension of space-time by two units through
\bq
 {\bf D}^- I_{\nu_1 \nu_2 \nu_3 \nu_4 \nu_5 \nu_6 \nu_7 \nu_8 \nu_9}\left( D \right)
 & = &
 I_{\nu_1 \nu_2 \nu_3 \nu_4 \nu_5 \nu_6 \nu_7 \nu_8 \nu_9}\left( D-2 \right).
\eq
We consider the sector $215$ with the propagators $P_1, P_2, P_3, P_5, P_7, P_8$.
\begin{figure}
\begin{center}
\includegraphics[scale=0.8]{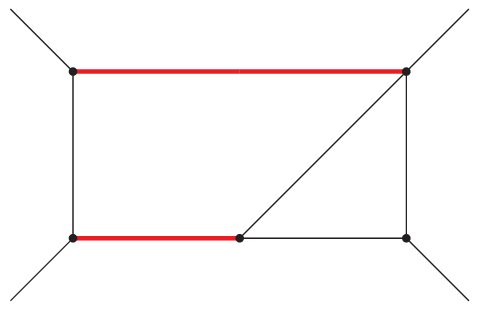}
\end{center}
\caption{
The Feynman diagram for the sector $215$. Black lines denote massless particles, red lines denote particles with a
mass $m$.
}
\label{fig_sector_215}
\end{figure}
The Feynman diagram for this sector is shown in fig.~\ref{fig_sector_215}.
The system has $16$ master integrals.
It is the minimal example with non-simultaneously rationalisable square roots.
A pre-canonical basis is given by
\bq
 & &
 I_{1 1 1 0 0 0 0 0 0},
 \;
 I_{1 0 1 0 0 0 1 0 0},
 \;
 I_{1 \left(-1\right) 1 0 0 0 1 0 0},
 \;
 I_{0 0 1 0 1 0 0 1 0},
 \;
 I_{1 1 1 0 0 0 1 0 0},
 \;
 I_{0 1 1 0 1 0 0 1 0},
 \;
 I_{0 1 1 0 0 0 1 1 0},
 \;
 I_{\left(-1\right) 1 1 0 0 0 1 1 0},
 \nonumber \\
 & &
 I_{1 1 1 0 1 0 1 0 0},
 \;
 I_{1 1 1 0 1 0 0 1 0},
 \;
 I_{1 1 1 0 0 0 1 1 0},
 \;
 I_{1 0 1 0 1 0 1 1 0},
 \;
 I_{1 \left(-1\right) 1 0 1 0 1 1 0},
 \;
 I_{0 1 1 0 1 0 1 1 0},
 \;
 I_{\left(-1\right) 1 1 0 1 0 1 1 0},
 \;
 I_{1 1 1 0 1 0 1 1 0}.
 \nonumber \\
\eq
An overview of the master integrals is given in table~\ref{table_master_integrals}.
\begin{table}
\begin{center}
\begin{tabular}{|c|r|r|l|l|c|}
\hline
 number of   & block & sector & master integrals & master integrals & roots \\
 propagators &       &        & basis $\vec{I}$  & basis $\vec{J}$  & \\
\hline
\hline
 $3$ & $1$ & $7$ & $I_{1}$ & $J_{1}$ & \\
     & $2$ & $69$ & $I_{2}, I_{3}$ & $J_{2}, J_{3}$ & \\
     & $3$ & $148$ & $I_{4}$ & $J_{4}$ & \\
\hline
 $4$ & $4$ & $71$ & $I_{5}$ & $J_{5}$ & $r_1$ \\
     & $5$ & $150$ & $I_{6}$ & $J_{6}$ & \\
     & $6$ & $198$ & $I_{7}, I_{8}$ & $J_{7}, J_{8}$ & $r_1$ \\
\hline
 $5$ & $7$ & $87$ & $I_{9}$ & $J_{9}$ & \\
     & $8$ & $151$ & $I_{10}$ & $J_{10}$ & \\
     & $9$ & $199$ & $I_{11}$ & $J_{11}$ & \\
     & $10$ & $213$ & $I_{12}, I_{13}$ & $J_{12}, J_{13}$ & \\
     & $11$ & $214$ & $I_{14}, I_{15}$ & $J_{14}, J_{15}$ & $r_2$ \\
\hline
 $6$ & $12$ & $215$ & $I_{16}$ & $J_{16}$ & $r_3$ \\
\hline
\end{tabular}
\end{center}
\caption{Overview of the set of master integrals.
The first column denotes the number of propagators, the second column labels consecutively the sectors,
the third column gives the sector id (defined in eq.~(\ref{def_sector_id})),
the fourth column lists the master integrals in the basis $\vec{I}$,
the fifth column the corresponding ones in the basis $\vec{J}$.
The last column denotes the dependence on square roots. 
}
\label{table_master_integrals}
\end{table}
There are four sectors (sectors $69$, $198$, $213$ and $214$) with two master integrals.
We may transform to a basis $I=(I_1,\dots,I_{16})^T$, which puts the differential equation into an
$\eps$-factorised form at the expense of introducing three square roots (see ref.~\cite{Heller:2019gkq}):
\bq
 r_1 
 & = & 
 \sqrt{ -s \left(4m^2-s\right)},
 \nonumber \\
 r_2 
 & = & 
 \sqrt{ \left(-s\right)\left(-t\right) \left[ 4m^2 \left(-t-m^2\right) -s \left(-t\right) \right]},
 \nonumber \\
 r_3
 & = &
 \sqrt{ \left(-s\right) \left[ \left(4m^2-s\right)\left(-t\right)^2 + 2 m^2 \left(-s\right)\left(-t\right) + m^4 \left(-s\right)\right]}.
\eq
Such a basis is given by
\begin{alignat}{2}
\label{def_masters_I}
 \mbox{Sector 7:} \;\;\;\; &
 I_{1}
 & = \;\; & 
 \eps^2 \frac{m^2}{\mu^2}
 \; {\bf D}^- I_{111000000},
 \nonumber \\
 \mbox{Sector 69:} \;\;\;\; &
 I_{2}
 & = \;\; & 
 \eps^2 \frac{\left(m^2-s\right)}{\mu^2}
 \; {\bf D}^- I_{101000100},
 \nonumber \\
 &
 I_{3}
 & = \;\; & 
 \eps^2 
 \left( {\bf D}^- I_{1\left(-1\right)1000100} - \frac{m^2}{\mu^2} \; {\bf D}^- I_{101000100} \right),
 \nonumber \\
 \mbox{Sector 148:} \;\;\;\; &
 I_{4}
 & = \;\; & 
 \eps^2 \frac{t}{\mu^2}
 \; {\bf D}^- I_{001010010},
 \nonumber \\
 \mbox{Sector 71:} \;\;\;\; &
 I_{5}
 & = \;\; & 
 \eps^2 \frac{r_1}{\mu^2}
 \left( \frac{m^2}{\mu^2} \; {\bf D}^- I_{111000100} - {\bf D}^- I_{101000100} \right), 
 \nonumber \\
 \mbox{Sector 150:} \;\;\;\; &
 I_{6}
 & = \;\; &
 \eps^3 \frac{t}{\mu^2}
 I_{012010010},
 \nonumber \\
 \mbox{Sector 198:} \;\;\;\; &
 I_{7}
 & = \;\; & 
 \eps^3 \frac{s}{\mu^2}
 I_{012000110},
 \nonumber \\
 &
 I_{8}
 & = \;\; & 
 r_1 \left( 
 \frac{1}{\eps} \frac{\partial}{\partial s} 
 -\frac{1}{s} \right) I_{7},
 \nonumber \\
 \mbox{Sector 87:} \;\;\;\; &
 I_{9}
 & = \;\; & 
 \eps^3 \frac{m^2 s}{\mu^4} 
 I_{211010100},
 \nonumber \\
 \mbox{Sector 151:} \;\;\;\; &
 I_{10}
 & = \;\; & 
 \eps^4 \frac{t}{\mu^2}
 I_{111010010},
 \nonumber \\
 \mbox{Sector 199:} \;\;\;\; &
 I_{11}
 & = \;\; & 
 \eps^4 \frac{s}{\mu^2}
 I_{111000110},
 \nonumber \\
 \mbox{Sector 213:} \;\;\;\; &
 I_{12}
 & = \;\; & 
 \eps^4 \frac{\left(s+t\right)}{\mu^2}
 I_{101010110},
 \nonumber \\
 &
 I_{13}
 & = \;\; & 
 \eps^3 \frac{m^2\left(s+t\right)}{\mu^4}
 I_{101010210},
 \nonumber \\
 \mbox{Sector 214:} \;\;\;\; &
 I_{14}
 & = \;\; & 
 \eps^3 \left(1-2\eps\right) \frac{s}{\mu^2}
 I_{011010110},
 \nonumber \\
 &
 I_{15}
 & = \;\; & 
 \eps^3 \frac{r_2}{\mu^4}
 I_{012010110},
 \nonumber \\
 \mbox{Sector 215:} \;\;\;\; &
 I_{16}
 & = \;\; & 
 \eps^4 \frac{r_3}{\mu^4} 
 I_{111010110}.
\end{alignat}
In this basis we have an $\eps$-factorised differential equation
\bq
 d I & = & \eps \tilde{A} I.
\eq
We write the differential one-forms appearing in $A$ as
\bq
\label{def_omega_j_part_1}
 \omega_j & = & d \ln l_j.
\eq
We call the $l_j$'s letters. In total there are $17$ letters\footnote{Setting for example $\mu=m$ reduces the number of letters by one. In this case we have $d\ln l_1=0$.}, 
which can be divided into rational letters and non-rational letters.
The rational letters are
\begin{align}
\label{def_omega_j_part_2}
 l_1 & = \frac{m^2}{\mu^2},
 &
 l_2 & = \frac{-s}{\mu^2},
 &
 l_3 & = \frac{-t}{\mu^2},
 \nonumber \\
 l_4 & = \frac{m^2-s}{\mu^2},
 &
 l_5 & = \frac{4m^2-s}{\mu^2},
 &
 l_6 & = \frac{m^2+t}{\mu^2},
 \nonumber \\
 l_7 & = \frac{-s-t}{\mu^2},
 &
 l_8 & = \frac{4m^2(-t-m^2)+st}{\mu^4},
 &
 l_9 & = \frac{4m^2t^2-s\left(m^2-t\right)^2}{\mu^6}.
\end{align}
The non-rational letters are
\begin{align}
\label{def_omega_j_part_3}
 l_{10} & = \frac{2m^2-s-r_1}{2m^2-s+r_1},
 &
 l_{11} & =\frac{\left(2m^2-s\right)\left(-t\right)-r_2}{\left(2m^2-s\right)\left(-t\right)+r_2},
 \nonumber \\
 l_{12} & = \frac{\left(-s\right)\left(-t\right)-r_2}{\left(-s\right)\left(-t\right)+r_2},
 &
 l_{13} & = \frac{\left(-s\right)\left[\left(4m^2-s\right)\left(-t\right)-2m^4\right]-r_1r_2}{\left(-s\right)\left[\left(4m^2-s\right)\left(-t\right)-2m^4\right]+r_1r_2},
 \nonumber \\
 l_{14} & = \frac{\left(-s\right)\left(m^2-t\right)-r_3}{\left(-s\right)\left(m^2-t\right)+r_3},
 &
 l_{15} & = \frac{\left(-s\right)\left(m^2+t\right)-r_3}{\left(-s\right)\left(m^2+t\right)+r_3},
 \nonumber \\
 l_{16} & = \frac{\left(-s\right)\left(st-m^2s-4m^2t\right)-r_1r_3}{\left(-s\right)\left(st-m^2s-4m^2t\right)+r_1r_3},
 &
 l_{17} & = \frac{s t \left(st-m^2s-4m^2t-2m^4\right)-r_2r_3}{s t \left(st-m^2s-4m^2t-2m^4\right)+r_2r_3}.
\end{align}
In the basis $I$ the diagonal blocks $\tilde{D}_i$ and the non-diagonal blocks $\tilde{N}_{ij}$ do not have any particular 
structure.
We give three examples: The diagonal block for sector $198$ is given by
\bq
 \tilde{D}_6
 & = &
 \left( \begin{array}{cc}
  -3\omega_1 + \omega_2 & \frac{1}{2} \omega_{10} \\
 \frac{3}{2} \omega_{10} & - \omega_1 - \omega_5 \\
 \end{array} \right),
\eq
the diagonal block for sector $214$ is given by
\bq
 \tilde{D}_{11}
 & = &
 \left( \begin{array}{cc}
 -3 \omega_1+2\omega_2-\frac{1}{2} \omega_3-\frac{1}{2} \omega_7 & \frac{1}{2} \omega_{11} \\
 - \frac{3}{2} \omega_{11} & \omega_1 - \frac{1}{2} \omega_3 + \frac{3}{2} \omega_7 - 2 \omega_8 \\
 \end{array} \right),
\eq
and the non-diagonal block $\tilde{N}_{(11)6}$ is given by
\bq
 \tilde{N}_{(11)6}
 & = &
 \left( \begin{array}{cc}
 - \frac{1}{2} \omega_3 + \frac{1}{2} \omega_7 & - \frac{1}{2} \omega_{10} \\
 \frac{3}{2} \omega_{11} & - \frac{1}{2} \omega_{13} \\
 \end{array} \right).
\eq

\subsubsection{A refined basis and the Galois group}

For the Drell--Yan example we may achieve the symmetries of eq.~(\ref{symmetry_1}), eq.~(\ref{symmetry_2}), eq.~(\ref{symmetry_3}) and
eq.~(\ref{symmetry_4}) for all sectors by a change to a new basis
$J=(J_1,\dots,J_{16})^T$ defined by
\begin{align}
\label{def_masters_J}
 & \mbox{Sector 69:} &
 J_2 & = I_2 + \sqrt{3} I_3,
 &
 J_3 & = I_2 - \sqrt{3} I_3,
 \nonumber \\
 & \mbox{Sector 198:} &
 J_7 & = I_7 + \frac{i}{3} \sqrt{3} I_8,
 &
 J_8 & = I_7 - \frac{i}{3} \sqrt{3} I_8,
 \nonumber \\
 & \mbox{Sector 87:} &
 J_9 & = I_9 - \frac{1}{8} \left( I_2 + I_3 \right),
 & &
 \nonumber \\
 & \mbox{Sector 199:} &
 J_{11} & = I_{11} + \frac{1}{2} I_{7},
 & &
 \nonumber \\
 & \mbox{Sector 213:} &
 J_{12} & = \sqrt{3} I_{12} - \left(1+\sqrt{3}\right) I_{13},
 &
 J_{13} & = -\sqrt{3} I_{12} - \left(1-\sqrt{3}\right) I_{13},
 \nonumber \\
 & \mbox{Sector 214:} &
 J_{14} & = 2 I_{14} + \frac{2}{3} \sqrt{3} I_{15},
 &
 J_{15} & = 2 I_{14} - \frac{2}{3} \sqrt{3} I_{15},
\end{align}
and $J_k=I_k$ for all other master integrals.
Apart from the three square roots $r_1, r_2, r_3$ we have two new (trivial) square roots
\bq
 s_3 \; = \; \sqrt{3},
 & &
 s_{-3} \; = \; i \sqrt{3}.
\eq
The square roots $s_3$ and $s_{-3}$ do not depend on the kinematic variables.
The Galois group is
\bq
 G & = & {\mathbb Z}_2 \times {\mathbb Z}_2 \times {\mathbb Z}_2 \times {\mathbb Z}_2 \times {\mathbb Z}_2,
\eq
with generators $\rho_1, \rho_2, \rho_3, \sigma_3, \sigma_{-3}$.
We adopt the convention that we label square roots, which depend on kinematic variables by $r_i$,
and square roots, which do not depend on kinematic variables by $s_i$.
The corresponding generators of the Galois group are denoted by $\rho_i$ and $\sigma_i$, respectively.
A generator $\rho_i$ acts on $r_i$ as
\bq
 \rho_i\left(r_i\right) & = & - r_i,
\eq
and trivially on all other square roots
\bq
 \rho_i\left(r_j\right) \; = \; r_j
 \;\;\;\;\;\;
 \mbox{for} \; i \neq j
 & \mbox{and} &
 \rho_i\left(s_k\right) \; = \; s_k.
\eq
The action of $\sigma_i$ is analogous:
\bq
 \sigma_i\left(s_i\right) & = & - s_i
\eq
and trivially on all other square roots.
In this new basis we have again an $\eps$-factorised differential equation
\bq
\label{diff_eq_J}
 d J & = & \eps A J.
\eq
The individual blocks are now symmetric, for example
the diagonal block for sector $198$ is now given by
\bq
 D_6
 & = &
 \left( \begin{array}{cc}
 -2\omega_1 + \frac{1}{2} \omega_2 - \frac{1}{2} \omega_5 & -\omega_1 + \frac{1}{2} \omega_2 + \frac{1}{2} \omega_5 + \frac{i\sqrt{3}}{2} \omega_{10} \\
 -\omega_1 + \frac{1}{2} \omega_2 + \frac{1}{2} \omega_5 - \frac{i\sqrt{3}}{2} \omega_{10} & -2\omega_1 + \frac{1}{2} \omega_2 - \frac{1}{2} \omega_5 \\
 \end{array} \right),
\eq
We have
\bq
 d^{(6)}_{11}
 & = &
 \rho_1\left(d^{(6)}_{11}\right)
 \; = \;
 \sigma_{-3}\left(d^{(6)}_{11}\right)
 \; = \;
 \rho_1\left(d^{(6)}_{22}\right)
 \; = \;
 \sigma_{-3}\left(d^{(6)}_{22}\right)
 \; = \;
 d^{(6)}_{22},
 \nonumber \\
 d^{(6)}_{21}
 & = &
 \rho_1\left(d^{(6)}_{12}\right)
 \; = \;
 \sigma_{-3}\left(d^{(6)}_{12}\right),
\eq
since
\bq
 \rho_1\left(\omega_{10}\right) & = & - \omega_{10}.
\eq
For the diagonal block of sector $214$ we find
\bq
 D_{11}
 & = &
 \left( \begin{array}{cc}
 - \omega_1 + \omega_2 - \frac{1}{2} \omega_3 + \frac{1}{2} \omega_7 - \omega_8 & -2\omega_1+\omega_2-\omega_7+\omega_8- \frac{\sqrt{3}}{2} \omega_{11} \\ 
 -2\omega_1+\omega_2-\omega_7+\omega_8+ \frac{\sqrt{3}}{2} \omega_{11} & - \omega_1 + \omega_2 - \frac{1}{2} \omega_3 + \frac{1}{2} \omega_7 - \omega_8 \\
 \end{array} \right).
\eq
We have (with $\rho_2(\omega_{11})=-\omega_{11}$)
\bq
 d^{(11)}_{11}
 & = &
 \rho_2\left(d^{(11)}_{11}\right)
 \; = \;
 \sigma_3\left(d^{(11)}_{11}\right)
 \; = \;
 \rho_2\left(d^{(11)}_{22}\right)
 \; = \;
 \sigma_3\left(d^{(11)}_{22}\right)
 \; = \;
 d^{(11)}_{22},
 \nonumber \\
 d^{(11)}_{21}
 & = &
 \rho_2\left(d^{(11)}_{12}\right)
 \; = \;
 \sigma_3\left(d^{(11)}_{12}\right).
\eq
For the non-diagonal block $N_{(11)6}$ we find
\bq
\lefteqn{
 N_{(11)6}
 = } & &
 \\
 & &
 \left( \begin{array}{cc}
 - \frac{1}{2} \omega_3 + \frac{1}{2} \omega_7 + \frac{i \sqrt{3}}{2} \omega_{10} + \frac{\sqrt{3}}{2} \omega_{11} + \frac{i}{2} \omega_{13} 
 &
 - \frac{1}{2} \omega_3 + \frac{1}{2} \omega_7 - \frac{i \sqrt{3}}{2} \omega_{10} + \frac{\sqrt{3}}{2} \omega_{11} - \frac{i}{2} \omega_{13} 
 \\
 - \frac{1}{2} \omega_3 + \frac{1}{2} \omega_7 + \frac{i \sqrt{3}}{2} \omega_{10} - \frac{\sqrt{3}}{2} \omega_{11} - \frac{i}{2} \omega_{13} 
 &
 - \frac{1}{2} \omega_3 + \frac{1}{2} \omega_7 - \frac{i \sqrt{3}}{2} \omega_{10} - \frac{\sqrt{3}}{2} \omega_{11} + \frac{i}{2} \omega_{13} 
 \end{array} \right).
 \nonumber
\eq
Here we have
\bq
 & &
 n^{((11)6)}_{12}
 = 
 \rho_1 \left( n^{((11)6)}_{11} \right)
 =  
 \sigma_{-3} \left( n^{((11)6)}_{11} \right),
 \\
 & &
 n^{((11)6)}_{21}
 = 
 \rho_2 \left( n^{((11)6)}_{11} \right)
 = 
 \sigma_3 \left( n^{((11)6)}_{11} \right),
 \nonumber \\
 & &
 n^{((11)6)}_{22}
 = 
 \rho_1 \rho_2 \left( n^{((11)6)}_{11} \right)
 = 
 \rho_1 \sigma_3 \left( n^{((11)6)}_{11} \right)
 = 
 \sigma_{-3} \rho_2 \left( n^{((11)6)}_{11} \right)
 = 
 \sigma_{-3} \sigma_3 \left( n^{((11)6)}_{11} \right).
\nonumber 
\eq
Note that 
\bq
 i & = & \frac{1}{3} s_3 s_{-3}
\eq
and therefore $\sigma_{-3}(i) = -i$.
The Galois symmetries associated with the four sectors with two master integrals each 
\begin{table}
\begin{center}
\begin{tabular}{|c|cccc|}
\hline
 Sector & $69$ & $198$ & $213$ & $214$ \\
\hline
 Galois symmetries & $\sigma_3$ & $\rho_1,\sigma_{-3}$ & $\sigma_3$ & $\rho_2,\sigma_3$ \\
\hline
\end{tabular}
\end{center}
\caption{Overview of the Galois symmetries of the sectors with more than one master integral.
}
\label{table_Galois_symmetries}
\end{table}
are summarised in table~\ref{table_Galois_symmetries}.
We have
\bq
 J_3
 & = & 
 \sigma_3\left(J_2\right),
 \nonumber \\
 J_8
 & = & 
 \rho_1\left(J_7\right)
 \; = \;
 \sigma_{-3}\left(J_7\right),
 \nonumber \\
 J_{13}
 & = & 
 \sigma_3\left(J_{12}\right),
 \nonumber \\
 J_{15}
 & = & 
 \rho_2\left(J_{14}\right)
 \; = \;
 \sigma_3\left(J_{14}\right).
\eq
Note that for the sectors $198$ and $214$ there is more than one element from the Galois group that relates the two
master integrals in this sector.
Furthermore, note that the occurrence of a square root in a particular sector does not imply
that this sector must have master integrals related by a Galois symmetry.
Counterexamples are given by sector $71$, where the definition of the master integral $J_5$ involves the square root
$r_1$ and by sector $215$, where the definition of the master integral $J_{16}$ involves the square root
$r_3$.
Both sectors have only one master integral. 

In summary, the matrix $A$ has the structure (for a compact notation we use hexadecimal indices)
{\small
\bq
\lefteqn{
 A
 = } & & \\
 & &
 \left( \begin{array}{cccccccccccccccc}
 a_{11} & 0 & 0 & 0 & 0 & 0 & 0 & 0 & 0 & 0 & 0 & 0 & 0 & 0 & 0 & 0 \\
 0 & \cellcolor{red} a_{22} & \cellcolor{orange} a_{23} & 0 & 0 & 0 & 0 & 0 & 0 & 0 & 0 & 0 & 0 & 0 & 0 & 0 \\
 0 & \cellcolor{orange} a_{32} & \cellcolor{red} a_{33} & 0 & 0 & 0 & 0 & 0 & 0 & 0 & 0 & 0 & 0 & 0 & 0 & 0 \\
 0 & 0 & 0 & a_{44} & 0 & 0 & 0 & 0 & 0 & 0 & 0 & 0 & 0 & 0 & 0 & 0 \\
 a_{51} & \cellcolor{pink} a_{52} & \cellcolor{pink} a_{53} & 0 & a_{55} & 0 & 0 & 0 & 0 & 0 & 0 & 0 & 0 & 0 & 0 & 0 \\
 a_{61} & 0 & 0 & a_{64} & 0 & a_{66} & 0 & 0 & 0 & 0 & 0 & 0 & 0 & 0 & 0 & 0 \\
 \cellcolor{olive} a_{71} & 0 & 0 & 0 & 0 & 0 & \cellcolor{green} a_{77} & \cellcolor{lime} a_{78} & 0 & 0 & 0 & 0 & 0 & 0 & 0 & 0 \\
 \cellcolor{olive} a_{81} & 0 & 0 & 0 & 0 & 0 & \cellcolor{lime} a_{87} & \cellcolor{green} a_{88} & 0 & 0 & 0 & 0 & 0 & 0 & 0 & 0 \\
 0 & 0 & 0 & 0 & a_{95} & 0 & 0 & 0 & a_{99} & 0 & 0 & 0 & 0 & 0 & 0 & 0 \\
 0 & 0 & 0 & 0 & 0 & a_{A6} & 0 & 0 & 0 & a_{AA} & 0 & 0 & 0 & 0 & 0 & 0 \\
 a_{B1} & \cellcolor{purple} a_{B2} & \cellcolor{purple} a_{B3} & 0 & a_{B5} & 0 & \cellcolor{teal} a_{B7} & \cellcolor{teal} a_{B8} & 0 & 0 & a_{BB} & 0 & 0 & 0 & 0 & 0 \\
 \cellcolor{Lavender} a_{C1} & 0 & \cellcolor{magenta} a_{C3} & \cellcolor{CornflowerBlue} a_{C4} & 0 & \cellcolor{CadetBlue} a_{C6} & 0 & 0 & 0 & 0 & 0 & \cellcolor{blue} a_{CC} & \cellcolor{cyan} a_{CD} & 0 & 0 & 0 \\
 \cellcolor{Lavender} a_{D1} & \cellcolor{magenta} a_{D2} & 0 & \cellcolor{CornflowerBlue} a_{D4} & 0 & \cellcolor{CadetBlue} a_{D6} & 0 & 0 & 0 & 0 & 0 & \cellcolor{cyan} a_{DC} & \cellcolor{blue} a_{DD} & 0 & 0 & 0 \\
 \cellcolor{BurntOrange} a_{E1} & 0 & 0 & \cellcolor{Dandelion} a_{E4} & 0 & \cellcolor{Goldenrod} a_{E6} & \cellcolor{yellow} a_{E7} & \cellcolor{yellow} a_{E8} & 0 & 0 & 0 & 0 & 0 & \cellcolor{gray} a_{EE} & \cellcolor{lightgray} a_{EF} & 0 \\
 \cellcolor{BurntOrange} a_{F1} & 0 & 0 & \cellcolor{Dandelion} a_{F4} & 0 & \cellcolor{Goldenrod} a_{F6} & \cellcolor{yellow} a_{F7} & \cellcolor{yellow} a_{F8} & 0 & 0 & 0 & 0 & 0 & \cellcolor{lightgray} a_{FE} & \cellcolor{gray} a_{FF} & 0 \\
 a_{01} & \cellcolor{violet} a_{02} & \cellcolor{violet} a_{03} & 0 & a_{05} & a_{06} & \cellcolor{GreenYellow} a_{07} & \cellcolor{GreenYellow} a_{08} & a_{09} & a_{0A} & a_{0B} & \cellcolor{Aquamarine} a_{0C} & \cellcolor{Aquamarine} a_{0D} & \cellcolor{brown} a_{0E} & \cellcolor{brown} a_{0F} & a_{00} \\
 \end{array} \right).
 \nonumber 
\eq
}
\noindent
Entries with the same background colour are related by a symmetry.

\subsubsection{The simplest example}

One of the simplest examples is given by the sector $69$ of the previous family of Feynman integrals.
It is worth discussing this example explicitly, as there is no square root with a dependence on the kinematic
variables associated to it.
This is the simplest example where the non-obvious square root $\sqrt{3}$ appears.
\begin{figure}
\begin{center}
\includegraphics[scale=0.8]{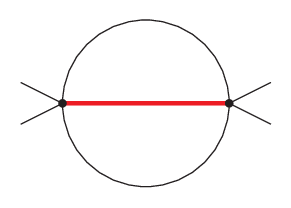}
\end{center}
\caption{
The Feynman diagram for the sector $69$. Black lines denote massless particles, red lines denote particles with a
mass $m$.
}
\label{fig_sector_69}
\end{figure}
The corresponding Feynman diagram is shown in fig.~\ref{fig_sector_69}.
This sector has no sub-sectors and forms a system with two master integrals, which we may takes as $(I_2,I_3)$,
with $I_2$ and $I_3$ defined in eq.~(\ref{def_masters_I}).
The differential equation for these master integrals reads
\bq
 d \left(\begin{array}{c} I_2 \\ I_3 \\ \end{array} \right)
 & = & 
 \eps \tilde{D}_2 \left(\begin{array}{c} I_2 \\ I_3 \\ \end{array} \right),
\eq
with
\bq
 \tilde{D}_2
 & = &
 \left( \begin{array}{cc}
  \frac{1}{2} \omega_1 + \frac{3}{2} \omega_2 - 4 \omega_4 & - \frac{3}{2} \omega_1 + \frac{3}{2} \omega_2 \\
 \frac{1}{2} \omega_1 - \frac{1}{2} \omega_2 & - \frac{3}{2} \omega_1 - \frac{1}{2} \omega_2 \\
 \end{array} \right).
\eq
The transformation from $(I_2,I_3)$ to $(J_2,J_3)$ given in eq.~(\ref{def_masters_J}) 
converts the differential equation to
\bq
 d \left(\begin{array}{c} J_2 \\ J_3 \\ \end{array} \right)
 & = & 
 \eps D_2 \left(\begin{array}{c} J_2 \\ J_3 \\ \end{array} \right)
\eq
with
\bq
 D_2
 =
 \left( \begin{array}{cc}
 - \frac{1}{2} \omega_1 + \frac{1}{2} \omega_2 - 2 \omega_4 & \left(1+\frac{\sqrt{3}}{2}\right) \omega_1 + \left(1-\frac{\sqrt{3}}{2}\right) \omega_2 - 2 \omega_4 \\
 \left(1-\frac{\sqrt{3}}{2}\right) \omega_1 + \left(1+\frac{\sqrt{3}}{2}\right) \omega_2 - 2 \omega_4 & - \frac{1}{2} \omega_1 + \frac{1}{2} \omega_2 - 2 \omega_4 \\
 \end{array} \right).
\eq
The self-duality requirement 
\bq
 d^{(2)}_{11} & = & d^{(2)}_{22}
\eq
will introduce the (non-obvious) square root $s_3=\sqrt{3}$.

\subsection{Electron-nucleon scattering}

A second interesting example 
\begin{figure}
\begin{center}
\includegraphics[scale=1.0]{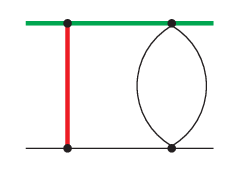}
\end{center}
\caption{
An example of a Feynman diagram relevant to electroweak corrections to electron-nucleon scattering.
Black lines denote massless particles, red lines denote particles with a
mass $m_Z$, green lines denote particles with a
mass $m_N$.
}
\label{fig_sector_55_topo_AC}
\end{figure}
is the family of Feynman diagrams shown in fig.~\ref{fig_sector_55_topo_AC},
contributing to the electroweak corrections to electron-nucleon scattering.
This example has one sector with two master integrals and two square roots.
The inverse propagators are now given by
\begin{align}
\label{def_inv_prop_mesa}
 P_1 & = -k_1^2,
 &
 P_2 & = -\left(k_1-p_1\right)^2 + m_Z^2,
 &
 P_3 & = -\left(k_1-p_{12}\right)^2 + m_N^2,
 \nonumber \\
 P_4 & = -\left(k_1-p_{123}\right)^2,
 &
 P_5 & = -k_2^2,
 &
 P_6 & = -\left(k_1+k_2-p_{123}\right)^2,
 \nonumber \\
 P_7 & = -\left(k_1+k_2+p_{12}\right)^2,
 & 
 P_8 & = -\left(k_1+k_2-p_{1}\right)^2,
 &
 P_9 & = -\left(k_1+k_2\right)^2,
\end{align}
and the external momenta satisfy
\bq
 p_1^2
 \; = \;
 p_4^2
 \; = \;
 0,
 & &
 p_2^2
 \; = \;
 p_3^2
 \; = \;
 m_N^2.
\eq
The Mandelstam variables $s$ and $t$ are defined as in eq.~(\ref{def_Mandelstam}).
In eq.~(\ref{def_inv_prop_mesa}) we used the short-hand notation $p_{12}=p_1+p_2$ and $p_{123}=p_{12}+p_3$.
A basis of master integrals, which puts the differential equation into an $\eps$-factorised form was given in ref.~\cite{Bottcher:2023wsr}
and reads
\begin{alignat}{2}
 \mbox{Sector 50:} \;\;\;\; &
 I_{1}
 & = \;\; & 
 \eps^2 
 \left(\frac{m_Z^2-t}{\mu^2}\right)
 \; {\bf D}^- I_{010011000},
 \nonumber \\
 &
 I_{2}
 & = \;\; & 
 \eps^2 
 \; {\bf D}^- I_{010\left(-1\right)11000},
 \nonumber \\
 \mbox{Sector 52:} \;\;\;\; &
 I_{3}
 & = \;\; & 
 \eps \left(1+4\eps\right) 
 \left( \frac{m_N^2}{\mu^2} \right)
 \; {\bf D}^- I_{001011000},
 \nonumber \\
 \mbox{Sector 53:} \;\;\;\; &
 I_{4}
 & = \;\; & 
 -4 \eps^3 
 \left(\frac{m_N^2-s}{\mu^2}\right)
 I_{101012000},
 \nonumber \\
 \mbox{Sector 54:} \;\;\;\; &
 I_{5}
 & = \;\; & 
 \eps^3 
 \left(\frac{r_1}{\mu^2}\right)
 I_{011012000},
 \nonumber \\
 &
 I_{6}
 & = \;\; & 
 \eps^2
 \left(\frac{r_3}{\mu^2}\right)
 \; {\bf D}^- I_{011\left(-1\right)11000},
 \nonumber \\
 \mbox{Sector 55:} \;\;\;\; &
 I_{7}
 & = \;\; & 
 \eps^3
 \left(\frac{m_N^2-s}{\mu^2}\right) \left(\frac{m_Z^2-t}{\mu^2}\right)
 I_{111012000},
 \nonumber \\
 &
 I_{8}
 & = \;\; & 
 \eps^3
 \left(\frac{m_N^2-s}{\mu^2}\right)
 I_{111\left(-1\right)12000}.
\end{alignat}
There are three sectors with two master integrals (sectors $50$, $54$ and $55$).
Sector $50$ in this example is an integral we encountered previously, it is the integral 
shown in fig.~\ref{fig_sector_69}.
Sector $54$ introduces two square roots $r_1$ and $r_3$, which are given by
\bq
\label{def_square_roots_topo_C}
 r_1 \; = \; \sqrt{-t\left(4m_N^2-t\right)},
 & &
 r_3 \; = \; \sqrt{-m_Z^2\left(4m_N^2-m_Z^2\right)}.
\eq
The occurrence of two square roots within one sector makes this example interesting.
Changing the basis of master integrals to
\begin{align}
\label{def_masters_J_topo_C}
 & \mbox{Sector 50:} &
 J_1 & = I_1 + \sqrt{3} I_2,
 &
 J_2 & = I_1 - \sqrt{3} I_2,
 \nonumber \\
 & \mbox{Sector 52:} &
 J_3 & = I_3,
 & &
 \nonumber \\
 & \mbox{Sector 53:} &
 J_4 & = I_4,
 & &
 \nonumber \\
 & \mbox{Sector 54:} &
 J_5 & = I_5 + \frac{i}{6} \sqrt{3} I_6,
 &
 J_6 & = I_5 - \frac{i}{6} \sqrt{3} I_6,
 \nonumber \\
 & \mbox{Sector 55:} &
 J_7 & = I_7 + \sqrt{3} I_8,
 &
 J_8 & = I_7 - \sqrt{3} I_8,
\end{align}
and $J_k=I_k$ for all other master integrals
will realise self-duality and the Galois symmetries.
In eq.~(\ref{def_masters_J_topo_C}) we introduced again two square roots
\bq
 s_3 \; = \; \sqrt{3},
 & &
 s_{-3} \; = \; i \sqrt{3}
\eq
with no dependence on the kinematic variables.
\begin{table}
\begin{center}
\begin{tabular}{|c|ccc|}
\hline
 Sector & $50$ & $54$ & $55$ \\
\hline
 Galois symmetries & $\sigma_3$ & $\rho_3,\sigma_{-3}$ & $\sigma_3$ \\
\hline
\end{tabular}
\end{center}
\caption{Overview of the Galois symmetries in the basis $J$ of the sectors with more than one master integral for the example shown in fig.~\ref{fig_sector_55_topo_AC}.
}
\label{table_Galois_symmetries_topo_C}
\end{table}
The Galois symmetries are summarised in table~\ref{table_Galois_symmetries_topo_C}.
We have
\bq
 J_2
 & = & 
 \sigma_3\left(J_1\right),
 \nonumber \\
 J_6
 & = & 
 \rho_3\left(J_5\right)
 \; = \;
 \sigma_{-3}\left(J_5\right),
 \nonumber \\
 J_8
 & = & 
 \sigma_3\left(J_7\right).
\eq
In the basis $J$ we have an $\eps$-factorised differential equation
\bq
 d J & = & \eps A J
\eq
where $A$ has the structure
\bq
\label{structure_A_mesabox}
 A & = &
 \left( \begin{array}{cccccccc}
 \cellcolor{red} a_{11} & \cellcolor{orange} a_{12} & 0 & 0 & 0 & 0 & 0 & 0 \\
 \cellcolor{orange} a_{21} & \cellcolor{red} a_{22} & 0 & 0 & 0 & 0 & 0 & 0 \\
 0 & 0 & a_{33} & 0 & 0 & 0 & 0 & 0 \\
 0 & 0 & a_{43} & a_{44} & 0 & 0 & 0 & 0 \\
 \cellcolor{yellow} a_{51} & \cellcolor{yellow} a_{52} & \cellcolor{ForestGreen} a_{53} & 0 & \cellcolor{green} a_{55} & \cellcolor{lime} a_{56} & 0 & 0 \\
 \cellcolor{yellow} a_{61} & \cellcolor{yellow} a_{62} & \cellcolor{ForestGreen} a_{63} & 0 & \cellcolor{lime} a_{65} & \cellcolor{green} a_{66} & 0 & 0 \\
 \cellcolor{magenta} a_{71} & \cellcolor{violet} a_{72} & \cellcolor{NavyBlue} a_{73} & \cellcolor{CornflowerBlue} a_{74} & \cellcolor{teal} a_{75} & \cellcolor{teal} a_{76} & \cellcolor{blue} a_{77} & \cellcolor{cyan} a_{78} \\
 \cellcolor{violet} a_{81} & \cellcolor{magenta} a_{82} & \cellcolor{NavyBlue} a_{83} & \cellcolor{CornflowerBlue} a_{84} & \cellcolor{teal} a_{85} & \cellcolor{teal} a_{86} & \cellcolor{cyan} a_{87} & \cellcolor{blue} a_{88} \\
 \end{array} \right).
\eq
Entries with the same background colour are related by a symmetry.

The basis of master integrals that maximises the symmetries of the matrix $A$
is not necessarily unique.
To see this, let us discuss the roles of the roots $r_1$ and $r_3$ in this example.
They appear in the same sector (sector $54$ with master integrals $J_5$ and $J_6$).
We have chosen a basis such that
\bq
 \rho_3\left(J_5\right) & = & J_6. 
\eq
Acting with $\rho_1$ on $J_5$ gives us
\bq
 \rho_1\left(J_5\right) & = & -J_6. 
\eq
Setting
\bq
 J_5' & = & I_6 - 2 i \sqrt{3} I_5,
 \nonumber \\
 J_6' & = & I_6 + 2 i \sqrt{3} I_5
\eq
and $J_k'=J_k$ for all other integrals will reverse the roles of $r_1$ and $r_3$.
We now have
\bq
 \rho_1\left(J_5'\right) \; = \; J_6',
 & &
 \rho_3\left(J_5'\right) \; = \; -J_6'.
\eq
In the basis $J'$ we have again an $\eps$-factorised differential equation
\bq 
 d J' 
 & = &
 \eps A' J'.
\eq
$A'$ has the same structure as the one shown in eq.~(\ref{structure_A_mesabox}).
\begin{table}
\begin{center}
\begin{tabular}{|c|ccc|}
\hline
 Sector & $50$ & $54$ & $55$ \\
\hline
 Galois symmetries & $\sigma_3$ & $\rho_1,\sigma_{-3}$ & $\sigma_3$ \\
\hline
\end{tabular}
\end{center}
\caption{Overview of the Galois symmetries in the basis $J'$ of the sectors with more than one master integral for the example shown in fig.~\ref{fig_sector_55_topo_AC}.
}
\label{table_Galois_symmetries_topo_C_Jprime}
\end{table}
The Galois symmetries in the basis $J'$ are summarised in table~\ref{table_Galois_symmetries_topo_C_Jprime}.

\subsection{An example with three master integrals within one sector}
\label{sect:example_three_masters}

As an example with a sector with three master integrals 
\begin{figure}
\begin{center}
\includegraphics[scale=1.0]{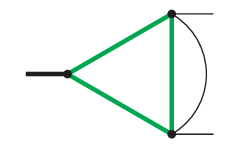}
\end{center}
\caption{
An example of a Feynman diagram contributing to the Higgs decay $H\rightarrow b \bar{b}$.
Thin black lines denote massless particles, a thick black line denotes a particle with a mass $m_H$,
green lines denote particles with a
mass $m_t$.
}
\label{fig_Hbb}
\end{figure}
we discuss a two-loop integral contributing to the Higgs decay $H \rightarrow b \bar{b}$, where the $b$-quarks are assumed to be massless.
The diagram is shown in fig.~\ref{fig_Hbb}.
This family has in total five master integrals, grouped into three sectors.
The first two sectors have just one master integral each.
The first sector (with one master integral) is given by a product of two one-loop tadpole integrals,
the second sector (again with one master integral) is given  by the product of a one-loop two-point function
with a one-loop tadpole integral.
The third sector has three master integrals.
We follow the notation of ref.~\cite{Chaubey:2019lum}.
The family of Feynman integrals has seven propagators
\begin{align}
 P_1 & = -k_1^2 + m_t^2,
 &
 P_2 & = -\left(k_1-p_1-p_2\right)^2 + m_t^2,
 &
 P_3 & = -\left(k_1+k_2\right)^2,
 \nonumber \\
 P_4 & = -\left(k_1+k_2-p_1\right)^2,
 &
 P_5 & = -k_2^2 + m_W^2,
 &
 P_6 & = -\left(k_2+p_2\right)^2 + m_t^2,
 \nonumber \\
 P_7 & = -\left(k_1-p_1\right)^2 + m_t^2.
 & &
\end{align}
We set $s=(p_1+p_2)^2$.
There is one square root
\bq
 r_1 & = & \sqrt{-s\left(4m_t^2-s\right)}.
\eq
A basis of master integrals, which puts the differential equation in $\eps$-factorised form is given 
by\footnote{In ref.~\cite{Chaubey:2019lum} these master integrals are denoted $J_1$, $J_3$, $J_{17}$, $J_{18}$ 
and $J_{19}$.}
\bq
 I_{1}
 & = &
 \eps^2 \; {\bf D}^- I_{1000010},
 \nonumber \\
 I_{2}
 & = &
 \frac{1}{2} \eps^2 \frac{r_1}{\mu^2} \; {\bf D}^- I_{1100010},
 \nonumber \\
 I_{3}
 & = &
 \eps^3 \frac{s}{\mu^2} \; I_{1120010},
 \nonumber \\
 I_{4}
 & = &
 \eps^3 \frac{s}{\mu^2} \; I_{1110020},
 \nonumber \\
 I_{5}
 & = &
 \eps^2 \frac{r_1}{\mu^2} \left[
 \left(1-2\eps\right) \; I_{2110010} + \eps \; I_{1110020} \right].
\eq
The change of basis $J_1=I_1$, $J_2=I_2$ and
\bq
 J_3 
 & = &
 I_3 + I_4 + \frac{i \sqrt{3}}{3} I_5,
 \nonumber \\
 J_4 
 & = &
 \frac{i \sqrt{6}}{3} I_3 + \frac{2 i \sqrt{6}}{3} I_4,
 \nonumber \\
 J_5 
 & = &
 I_3 + I_4 - \frac{i \sqrt{3}}{3} I_5,
\eq
introduces two constant square roots
\bq
 s_{-3} \; = \; i \sqrt{3},
 & &
 s_{-6} \; = \; i \sqrt{6},
\eq
and realises self-duality and the Galois symmetry.
The differential equation reads
\bq
 d J & = & \eps A J,
 \;\;\;
 A \; = \;
 \left( \begin{array}{ccccc}
 0     & 0 & 0 & 0 & 0 \\
 a_{21} & a_{22} & 0 & 0 & 0 \\
 \cellcolor{pink} a_{31} & 0     & \cellcolor{red} a_{33} & \cellcolor{green} a_{34} & \cellcolor{yellow} a_{35} \\
 0     & a_{42} & \cellcolor{green} a_{43} & \cellcolor{cyan} a_{44} & \cellcolor{green} a_{45} \\
 \cellcolor{pink} a_{51} & 0     & \cellcolor{yellow} a_{53} & \cellcolor{green} a_{54} & \cellcolor{red} a_{55} \\
 \end{array} \right).
\eq
Self-duality corresponds to the relations
\bq
 a_{55} \; = \; a_{33},
 \;\;\;\;\;\;
 a_{45} \; = \; a_{34},
 \;\;\;\;\;\;
 a_{54} \; = \; a_{43}.
\eq
The Galois symmetry $J_3=\rho_1(J_2)=\sigma_{-3}(J_2)$ gives in addition the relations
\bq
 & &
 a_{43} \; = \; \rho_1\left(a_{34}\right) \; = \; \sigma_{-3}\left(a_{34}\right),
 \nonumber \\
 & &
 a_{53} \; = \; \rho_1\left(a_{35}\right) \; = \; \sigma_{-3}\left(a_{35}\right),
 \nonumber \\
 & &
 a_{51} \; = \; \rho_1\left(a_{31}\right) \; = \; \sigma_{-3}\left(a_{31}\right).
\eq
It is worth noting that
\bq
 \sqrt{2} & = & - \frac{1}{3} s_{-3} s_{-6},
\eq
and therefore
\bq
\label{example_galois_sqrt2_Isqrt6}
 \sigma_{-3}\left(\sqrt{2}\right) \; = \; - \sqrt{2}
 & \mbox{and} &
 \sigma_{-3}\left(i \sqrt{6}\right) \; = \; i \sqrt{6}. 
\eq
The latter relation states that $\sigma_{-3}$ acts on $s_{-3}$, but not on $s_{-6}$. 
If we set $\mu=m_t$, the entries of $A'$ are linear combinations of
\bq
 \omega_1
 \; = \; 
 \frac{ds}{s},
 \;\;\;\;\;\;
 \omega_2
 \; = \;
 \frac{ds}{s-4m_t^2},
 \;\;\;\;\;\;
 \omega_3
 \; = \;
 \frac{ds}{\sqrt{-s\left(4m_t^2-s\right)}}
 \; = \;
 \frac{1}{2} d \ln \frac{2m_t^2-s-r_1}{2m_t^2-s+r_1}.
\eq
The differential one-form $\omega_3$ is odd under $\rho_1$:
\bq
\label{example_galois_omega1}
 \rho_1\left(\omega_3\right) & = & - \omega_3.
\eq
The entries $a_{34}$ and $a_{43}$ are given by
\bq
 a_{34} \; = \; - \frac{i\sqrt{6}}{2} \omega_1 + \sqrt{2} \omega_3,
 & &
 a_{43} \; = \; - \frac{i\sqrt{6}}{2} \omega_1 - \sqrt{2} \omega_3,
\eq
and with eq.~(\ref{example_galois_sqrt2_Isqrt6}) and eq.(\ref{example_galois_omega1}) one directly verifies the relation
$a_{43} = \rho_1(a_{34}) = \sigma_{-3}(a_{34})$.
The symmetries of the diagonal block corresponding to sector $39$ (the $(3 \times 3)$-block formed by the master integrals
$J_3$, $J_4$ and $J_5$) follow the pattern of symmetries shown in fig.~\ref{fig_symmetries}.

\subsection{An example with four master integrals within one sector}

As an advanced example we discuss a three-loop integral contributing to the Higgs boson self-energy.
\begin{figure}
\begin{center}
\includegraphics[scale=0.8]{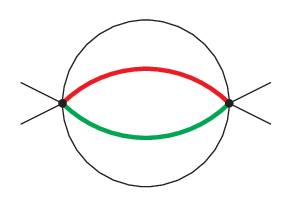}
\end{center}
\caption{
An example of a Feynman diagram contributing to the three-loop Higgs boson self-energy.
Black lines denote massless particles, red lines denote particles with a
mass $m_W$, green lines denote particles with a
mass $m_t$.
}
\label{fig_threeloop}
\end{figure}
The diagram is shown in fig.~\ref{fig_threeloop}.
This family has a single sector with four master integrals.
We follow the notation of ref.~\cite{Chaubey:2022jfi}.
We consider the integrals
\bq
 I_{\nu_1 \nu_2 \nu_3 \nu_4 \nu_5 \nu_6 \nu_7 \nu_8 \nu_9}
 = 
 e^{3 \gamma_E \eps}
 \left(\mu^2\right)^{\nu-\frac{3}{2}D}
 \int \left( \prod\limits_{a=1}^{3} \frac{d^Dk_a}{i \pi^{\frac{D}{2}}} \right)
 \left( \prod\limits_{c=1}^{9} \frac{1}{P_c^{\nu_c}} \right),
\eq
where the inverse propagators are given by
\begin{align}
 P_1
 & =
 -k_1^2 + m_t^2,
 &
 P_2
 & =
 -\left(k_1-p\right)^2 + m_t^2,
 &
 P_3
 & =
 -\left(k_1+k_2\right)^2,
 \nonumber \\
 P_4
 & =
 -k_2^2 + m_t^2,
 &
 P_5
 & =
 -\left(k_2+k_3\right)^2 + m_W^2,
 &
 P_6
 & =
 -\left(k_2+p\right)^2 + m_t^2,
 \nonumber \\
 P_7
 & =
 -k_3^2,
 & 
 P_8
 & =
 -\left(k_3-p\right)^2,
 &
 P_9
 & =
 -\left(k_1-k_3\right)^2 + m_t^2.
\end{align}
We are interested in sector $86$ (with propagators $2,3,5,7$).
A basis of master integrals that puts the differential equation into an $\eps$-factorised form is given by
\bq
 I_{1}
 & = &
 \eps^3 \; \frac{r_2}{\mu^2} \; {\bf D}^- I_{011010100},
 \nonumber \\
 I_{2}
 & = &
 \eps^3 \; \left[ {\bf D}^- I_{011\left(-1\right)10100} - \frac{\left(m_t^2-m_W^2\right)}{\mu^2} \; {\bf D}^- I_{011010100} \right],
 \nonumber \\
 I_{3}
 & = &
 \eps^3 \; {\bf D}^- I_{01101\left(-1\right)100},
 \nonumber \\
 I_{4}
 & = &
 \eps^3 \; \left[ {\bf D}^- I_{0110101\left(-1\right)0} + \frac{p^2}{\mu^2} \; {\bf D}^- I_{011010100} \right].
\eq
In this example we have the square root of the K\"allen function
\bq
 r_2
 & = &
 \sqrt{\lambda\left(p^2,m_W^2,m_t^2\right)}.
\eq
We recall that the K\"allen function is defined by
\bq
 \lambda(x,y,z)
 & = &
 x^2 + y^2 + z^2 - 2 x y - 2 y z - 2 z x.
\eq
The change of basis
\bq
\label{def_basis_four_masters}
 J_1
 & = &
 \frac{i\sqrt{2}}{2} I_1 + I_2 + I_3 + I_4,
 \nonumber \\
 J_2
 & = &
 \frac{1}{2} \left(i + \sqrt{3}\right) I_2 + \frac{1}{2} \left(i - \sqrt{3}\right) I_3 - i I_4, 
 \nonumber \\
 J_3
 & = &
 \frac{1}{2} \left(i - \sqrt{3}\right) I_2 + \frac{1}{2} \left(i + \sqrt{3}\right) I_3 - i I_4, 
 \nonumber \\
 J_4
 & = &
 - \frac{i\sqrt{2}}{2} I_1 + I_2 + I_3 + I_4,
\eq
introduces three additional square roots
\bq
 s_3 \; = \; \sqrt{3},
 \;\;\;\;\;\;
 s_{-2} \; = \; i \sqrt{2},
 \;\;\;\;\;\;
 s_{-1} \; = \; i.
\eq
From the definition of the basis $J$ it follows that
\bq
\label{Galois_symmetries_rho2_sigmam2_sigma3}
 J_4 & = & \rho_2\left(J_1\right) \; = \; \sigma_{-2}\left(J_1\right),
 \nonumber \\
 J_3 & = & \sigma_3\left(J_2\right).
\eq 
With the choice of the master integrals as in eq.~(\ref{def_basis_four_masters}) the square root $s_{-1}=i$ 
leads to the relation $J_3 =  - \sigma_{-1}(J_2)$. However, it is sufficient to focus on the Galois symmetries in eq.~(\ref{Galois_symmetries_rho2_sigmam2_sigma3}).
In the basis $J$, the differential equation is again in an $\eps$-factorised form
\bq
 d J & = & \eps A J,
\eq
and is structured as follows
\bq
 A
 & = &
 \left( \begin{array}{cccc}
 \cellcolor{red} a_{11} & \cellcolor{yellow} a_{12} & \cellcolor{yellow} a_{13} & \cellcolor{orange} a_{14} \\
 \cellcolor{yellow} a_{21} & \cellcolor{green} a_{22} & \cellcolor{lime} a_{23} & \cellcolor{yellow} a_{24} \\
 \cellcolor{yellow} a_{31} & \cellcolor{lime} a_{32} & \cellcolor{green} a_{33} & \cellcolor{yellow} a_{34} \\
 \cellcolor{orange} a_{41} & \cellcolor{yellow} a_{42} & \cellcolor{yellow} a_{43} & \cellcolor{red} a_{44} \\
 \end{array} \right).
\eq
We have the self-duality relations
\bq
 a_{44} \;= \; a_{11},
 \;\;\;\;\;\;
 a_{33} \;= \; a_{22},
 \;\;\;\;\;\;
 a_{34} \;= \; a_{12},
 \;\;\;\;\;\;
 a_{24} \;= \; a_{13},
 \;\;\;\;\;\;
 a_{42} \;= \; a_{31},
 \;\;\;\;\;\;
 a_{43} \;= \; a_{21}.
\eq
In addition we have the Galois symmetries
\bq
 a_{13} \; = \; \sigma_3\left(a_{12}\right),
 \;\;\;\;\;\;
 a_{31} \; = \; \sigma_3\left(a_{21}\right),
 \;\;\;\;\;\;
 a_{42} \; = \; \rho_2\left(a_{12}\right) \; = \; \sigma_{-2}\left(a_{12}\right).
\eq
We see that self-duality and Galois symmetries reduce the number of entries of $A$, which need to be known,
to five (for example $a_{11}$, $a_{22}$, $a_{14}$, $a_{23}$ and $a_{12}$), the remaining ones follow from
symmetry.

\subsection{The massless double-box integral}
\label{sect:example_doublebox}

\begin{figure}
\begin{center}
\includegraphics[scale=0.8]{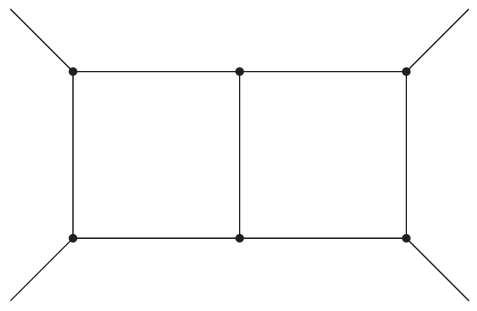}
\end{center}
\caption{
The Feynman diagram for the massless planar double-box integral.
}
\label{fig_double_box}
\end{figure}

While we have seen that in sectors with more than one master integral we may have in addition to self-duality a Galois symmetry, 
this is not guaranteed.
To illustrate this point, we discuss one of the simplest examples, the massless planar double-box integral \cite{Smirnov:1999gc} depicted in fig~\ref{fig_double_box}.
We show that in this case we have a limit Galois symmetry as discussed in section~\ref{sect:limit_Galois_symmetry}.
This example has eight master integrals, grouped into seven sectors.
The first six sectors have one master integral each, the top sector consists of two master integrals.
An $\eps$-factorised form has been given in \cite{Henn:2013pwa,Henn:2014qga}.
We follow the notation of ref.~\cite{Weinzierl:2022eaz}.
The inverse propagators are given by
\begin{align}
 P_1 & = -\left(k_1-p_1\right)^2,
 &
 P_2 & = -\left(k_1-p_1-p_2\right)^2,
 &
 P_3 & = -k_1^2,
 \nonumber \\
 P_4 & = -\left(k_1+k_2\right)^2,
 &
 P_5 & = -\left(k_2+p_1+p_2\right)^2,
 &
 P_6 & = -k_2^2,
 \nonumber \\
 P_7 & = -\left(k_2 + p_1 + p_2 + p_3\right)^2,
 & 
 P_8 & = -\left(k_1-p_1-p_3\right)^2,
 &
 P_9 & = -\left(k_2+p_1+p_3\right)^2.
\end{align}
We denote the relevant kinematic variable by $x=s/t$, with $s$ and $t$ being the usual Mandelstam variables as defined in eq.~\eqref{def_Mandelstam}. 
There are two differential one-forms,
\bq
 \omega_0 \; = \; d\ln\left(x\right) & \mbox{and} & \omega_1 \;= \; d\ln\left(x+1\right).
\eq 
Starting from the pre-canonical basis
\bq
\lefteqn{
 K = } & & \\
 & & \left(
 I_{001110000}, 
 I_{100100100},
 I_{011011000},
 I_{100111000},
 I_{111100100},
 I_{101110100},
 I_{111111100}, I_{1111111\left(-1\right)0} \right)^T
 \nonumber
\eq
we obtain an $\eps$-factorised basis
through
\bq
\label{def_basis_I_doublebox}
 I & = & U' U K,
\eq
where $U$ is given in eq.~(6.232) of ref.~\cite{Weinzierl:2022eaz}
and $U'$ is given by
\bq
 U'
 & = &
 \left( \begin{array}{rrrrrrrr}
1&0&0&0&0&0&0&0\\
0&1&0&0&0&0&0&0\\
0&0&1&0&0&0&0&0\\
0&0&0&1&0&0&0&0\\
0&0&0&1&1&0&0&0\\
0&-1&0&0&0&1&0&0\\
2&-3&-1&1&0&-2&1&0\\
-2&2&0&-2&0&4&0&1\\
 \end{array} \right).
\eq
The constant matrix $U'$ maximises zeros in the last two rows.
In the basis $I$
we have a differential equation in $\eps$-factorised form.
It is easily checked that it is impossible to have self-duality and a Galois symmetry at the same time.
There are no square roots which depend on the kinematic variable $x$.
We may make self-duality manifest through a $\mathrm{GL}(2,{\mathbb C})$-transformation in the top sector.
In this case we find that the required transformation is actually a $\mathrm{GL}(2,{\mathbb Q})$-transformation
and does not introduce any square roots either.
Hence, there is no Galois symmetry (in the usual sense) on top of self-duality.
However, there is a limit Galois symmetry.
The change of basis
\bq
 J_7 \; = \; I_7 + \left( 1+ \frac{s_1}{2} \right) I_8,
 & &
 J_8 \; = \; I_7 + \left( 1- \frac{s_1}{2} \right) I_8
\eq
with $s_1=\sqrt{1}$
makes self-duality manifest.
In addition we have a limit Galois symmetry
\bq
 J_8 & = & \sigma_1\left(J_7\right).
\eq
In practical terms we set $s_1=1$ in the end and the limit Galois symmetry is simply the substitution
$s_1 \rightarrow -s_1$.
In the basis $J$ we have the $\eps$-factorised differential equation
\bq
 d J & = & \eps A J
\eq
with
\bq
\label{eq_A_doublebox}
 A & = &
 \left( \begin{array}{cccccccc}
 a_{11} & 0 & 0 & 0 & 0 & 0 & 0 & 0 \\
 0 & 0 & 0 & 0 & 0 & 0 & 0 & 0 \\
 0 & 0 & a_{33} & 0 & 0 & 0 & 0 & 0 \\
 0 & 0 & 0 & a_{44} & 0 & 0 & 0 & 0 \\
 0 & a_{52} & 0 & a_{54} & a_{55} & 0 & 0 & 0 \\
 a_{61} & a_{62} & 0 & 0 & 0 & a_{66} & 0 & 0 \\
 0 & \cellcolor{BurntOrange} a_{72} & \cellcolor{Dandelion} a_{73} & 0 & \cellcolor{Goldenrod} a_{75} & \cellcolor{yellow} a_{76} & \cellcolor{red} a_{77} & \cellcolor{orange} a_{78} \\
 0 & \cellcolor{BurntOrange} a_{82} & \cellcolor{Dandelion} a_{83} & 0 & \cellcolor{Goldenrod} a_{85} & \cellcolor{yellow} a_{86} & \cellcolor{orange} a_{87} & \cellcolor{red} a_{88} \\
 \end{array} \right),
\eq
where entries with the same background colour are related by self-duality or a limit Galois symmetry.
In detail we have
\bq
 & &
 a_{77} \; = \; a_{88} \; = \; -2 \omega_0 + \frac{1}{2} \omega_1,
 \nonumber \\
 & &
 a_{78} \; = \; -2\left(1+s_1\right) \omega_0 + \frac{1}{2} \omega_1,
 \;\;\;\;\;\;
 a_{87} \; = \; -2\left(1-s_1\right) \omega_0 + \frac{1}{2} \omega_1.
\eq
The entries $a_{78}$ and $a_{87}$ are related by
\bq
 a_{87} & = & \sigma_1\left(a_{78}\right).
\eq
The entries of the non-diagonal blocks are given by
\bq
 a_{72} = a_{82} = -4 \omega_0,
 \;\;\;\;\;\;
 a_{73} = a_{83} = 2 \omega_0,
 \;\;\;\;\;\;
 a_{75} = a_{85} = -2 \omega_1,
 \;\;\;\;\;\;
 a_{76} = a_{86} = -4 \omega_1.
\eq

\subsection{The planar pentabox integral}
\label{sect:pentabox}

\begin{figure}
\begin{center}
\includegraphics[scale=0.7]{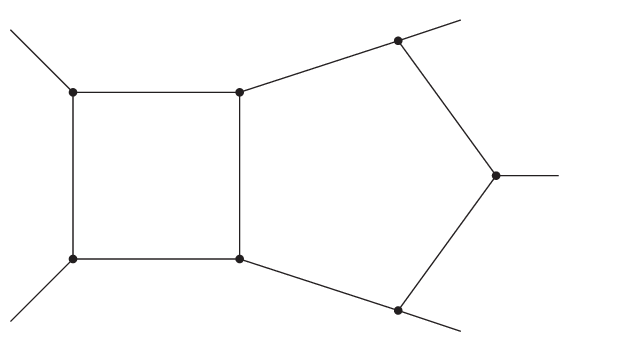}
\end{center}
\caption{
The Feynman diagram for the planar massless pentabox integral.
}
\label{fig_pentabox}
\end{figure}

As our most involved example we discuss the planar massless pentabox integral shown in fig.~\ref{fig_pentabox}.
This is an example with $61$ master integrals.
An $\eps$-factorised form has been given in ref.~\cite{Gehrmann:2018yef}.
The motivation for this example is as follows:
Up to now all examples where we had Galois symmetries on top of self-duality involved Feynman integrals with at least one massive internal propagator.
On the other hand, the massless double-box integral discussed in section~\ref{sect:example_doublebox}
had only a limit Galois symmetry on top of self-duality.
The planar massless pentabox integral is a Feynman integral that has in almost all sectors (normal) Galois symmetries in addition to self-duality
with the exception of two sectors, where we find limit Galois symmetries in addition self-duality.
The second aspect which makes this example interesting is the fact that in two sectors we find 
one-parameter families of Galois symmetries.

There is one kinematic square root
\bq
 r_1 & = & \sqrt{\Delta},
 \;\;\;\;\;\;\;\;\;
 \Delta \; = \; 
 \left( 4 i \eps_{\mu\nu\rho\sigma} p_1^\mu p_2^\nu p_3^\rho p_4^\sigma \right)^2.
\eq
The alphabet consists of $26$ letters in the planar case. 
$21$ letters are even with respect to $r_1 \rightarrow -r_1$, 
five letters are odd.
The $61$ master integrals group into $46$ sectors. $34$ sectors have one master integral, $9$ sectors have two master integrals
and $3$ sectors have three master integrals.
We find that it is possible to redefine the master integrals such that self-duality is manifest in all sectors.
Furthermore it is possible to choose the master integrals such that we have Galois symmetries
in all sectors except two.
In these two sectors we do have limit Galois symmetries.
The two sectors are the ones corresponding to the two double-box topologies with one external massive leg.

We denote by $I=(I_1,\dots,I_{61})^T$ the basis of master integrals as given in ref.~\cite{Gehrmann:2018yef}.
The symmetries are realised by the change of basis
\begin{align}
 J_{22} & = I_{22} + \frac{\sqrt{3}}{6} I_{23},
 &
 J_{23} & = I_{22} - \frac{\sqrt{3}}{6} I_{23},
 \nonumber \\
 J_{26} & = I_{26} + \frac{\sqrt{3}}{6} I_{27},
 &
 J_{27} & = I_{26} - \frac{\sqrt{3}}{6} I_{27},
 \nonumber \\
 J_{31} & = I_{31} + \frac{\sqrt{3}}{6} I_{32},
 &
 J_{32} & = I_{31} - \frac{\sqrt{3}}{6} I_{32},
 \nonumber \\
 J_{37} & = I_{38} + \frac{2i\sqrt{3}}{3} I_{37},
 &
 J_{38} & = I_{38} - \frac{2i\sqrt{3}}{3} I_{37},
 \nonumber \\
 J_{40} & = I_{40} + \frac{i\sqrt{3}}{3} I_{41},
 &
 J_{41} & = I_{40} - \frac{i\sqrt{3}}{3} I_{41},
 \nonumber \\
 J_{44} & = I_{44} + \frac{i\sqrt{3}}{3} I_{45},
 &
 J_{45} & = I_{44} - \frac{i\sqrt{3}}{3} I_{45},
 \nonumber \\
 J_{46} & = I_{46} + \frac{i\sqrt{3}}{3} I_{47},
 &
 J_{47} & = I_{46} - \frac{i\sqrt{3}}{3} I_{47},
 \nonumber \\
 J_{49} & = I_{49} + \frac{\left(3-\lambda_1\right)}{2} I_{50} + 2 \sqrt{\lambda_1} I_{51},
 &
 J_{51} & = I_{49} + \frac{\left(3-\lambda_1\right)}{2} I_{50} - 2 \sqrt{\lambda_1} I_{51},
 \nonumber \\
 & &
 J_{50} & = i \sqrt{2} I_{49} + \frac{i \sqrt{2}\left(3+\lambda_1\right)}{2} I_{50},
 \nonumber \\
 J_{52} & = I_{53} + \frac{1}{3} \left(1+s_1\right) I_{52},
 &
 J_{53} & = I_{53} + \frac{1}{3} \left(1-s_1\right) I_{52},
 \nonumber \\
 J_{54} & = I_{55} + \frac{1}{3} \left(1+s_1\right) I_{54},
 &
 J_{55} & = I_{55} + \frac{1}{3} \left(1-s_1\right) I_{54},
 \nonumber \\
 J_{56} & = I_{56} + \frac{\left(3-\lambda_2\right)}{2} I_{57} + 4 \sqrt{\lambda_2} I_{58},
 &
 J_{58} & = I_{56} + \frac{\left(3-\lambda_2\right)}{2} I_{57} - 4 \sqrt{\lambda_2} I_{58},
 \nonumber \\
 & &
 J_{57} & = i \sqrt{2} I_{56} + \frac{i \sqrt{2}\left(3+\lambda_2\right)}{2} I_{57},
 \nonumber \\
 J_{59} & = I_{60} + \frac{7 \sqrt{2}}{2} I_{59} + \frac{\sqrt{2}}{2} I_{61},
 &
 J_{61} & = I_{60} - \frac{7 \sqrt{2}}{2} I_{59} - \frac{\sqrt{2}}{2} I_{61},
 \nonumber \\
 & &
 J_{60} & =  5 I_{59} + 3 I_{61}.
\end{align}
and $J_k=I_k$ for all other master integrals.
Here we introduced the square roots
\bq
 s_1 \; = \; \sqrt{1}, 
 \;\;\;
 s_2 \; = \; \sqrt{2}, 
 \;\;\;
 s_3 \; = \; \sqrt{3}, 
 \;\;\;
 s_{-3} \; = \; i \sqrt{3},
 \;\;\;
 s_{\lambda_1} \; = \; \sqrt{\lambda_1},
 \;\;\;
 s_{\lambda_2} \; = \; \sqrt{\lambda_2}.
\eq
The Galois symmetries associated with the sectors consisting of two or three master integrals
\begin{table}
\begin{center}
\begin{tabular}{|c|c|}
\hline
 Sector & Galois symmetries\\
\hline
 $J_{22}, J_{23}$ & $\sigma_3$ \\
 $J_{26}, J_{27}$ & $\sigma_3$ \\
 $J_{31}, J_{32}$ & $\sigma_3$ \\
 $J_{37}, J_{38}$ & $\rho_1, \sigma_{-3}$ \\
 $J_{40}, J_{41}$ & $\rho_1, \sigma_{-3}$ \\
 $J_{44}, J_{45}$ & $\rho_1, \sigma_{-3}$ \\
 $J_{46}, J_{47}$ & $\rho_1, \sigma_{-3}$ \\
 $J_{49}, J_{50}, J_{51}$ & $\rho_1, \sigma_{\lambda_1}$ \\
 $J_{52}, J_{53}$ & $\sigma_1$ \\
 $J_{54}, J_{55}$ & $\sigma_1$ \\
 $J_{56}, J_{57}, J_{58}$ & $\rho_1, \sigma_{\lambda_2}$ \\
 $J_{59}, J_{60}, J_{61}$ & $\sigma_2$ \\
\hline
\end{tabular}
\end{center}
\caption{Overview of the Galois symmetries of the sectors with more than one master integral.
}
\label{table_Galois_symmetries_pentabox}
\end{table}
are summarised in table~\ref{table_Galois_symmetries_pentabox}.
The sectors $(J_{52}, J_{53})$ and
$(J_{54}, J_{55})$
have a limit Galois symmetry.
Of particular interest are also the sectors
$(J_{49}, J_{50}, J_{51})$ and $(J_{56}, J_{57}, J_{58})$, where we have a parameterised Galois symmetry:
For any value $\lambda_i \in {\mathbb Q}$ (with $i=1,2$) which is not a perfect square we 
have self-duality and a Galois symmetry.

We further remark that it is a matter of convention whether $\rho_1$ is considered to be a Galois
symmetry of the top sector $(J_{59}, J_{60}, J_{61})$.
This is a sector with three master integrals and we required in this paper that any Galois
symmetry acts trivially on the 
middle (second) master integral (see eq.~(\ref{Galois_action_middle_master_integral})).
In our example $\rho_1$ acts on $J_{60}$ as
\bq
 \rho_1\left(J_{60}\right) & = & - J_{60}.
\eq
If one takes this additional minus sign into account one may view $\rho_1$ also as a Galois symmetry
of the top sector.

\subsection{Further examples}

In addition we investigated several families of Feynman integrals with about ${\mathcal O}(30)$ master integrals each.
First, we considered the full set of the two-loop mixed electroweak-QCD master integrals for the Drell--Yan process
as discussed in refs.~\cite{Heller:2019gkq,vonManteuffel:2017myy,Bonciani:2016ypc}.
This family consists of $36$ master integrals.
\begin{figure}
\begin{center}
\includegraphics[scale=0.6]{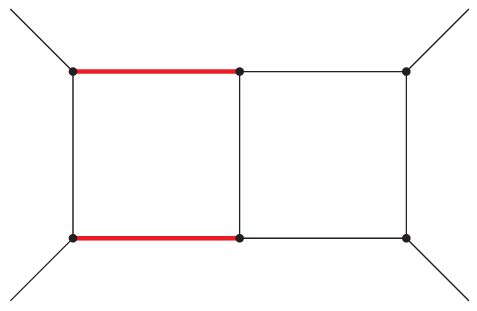}
\hspace*{15mm}
\includegraphics[scale=0.6]{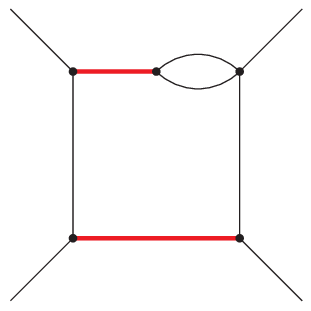}
\end{center}
\caption{
The top-level diagrams for the two-loop mixed electroweak-QCD master integrals for the Drell--Yan process.
Black lines denote massless particles, red lines denote particles with a mass $m$.
}
\label{fig_massive_double_box_drell_yan}
\end{figure}
The top-level diagrams are shown in fig.~\ref{fig_massive_double_box_drell_yan}.
This family contains the example discussed in section~\ref{sect:drell_yan} as a sub-system.
In addition, it has a sector with four master integrals.

Secondly, we investigated the family of the planar double-box integrals with one internal mass.
\begin{figure}
\begin{center}
\includegraphics[scale=0.6]{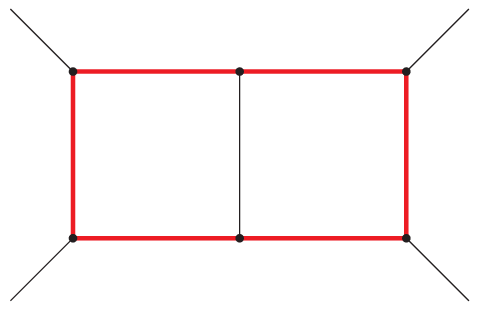}
\hspace*{15mm}
\includegraphics[scale=0.6]{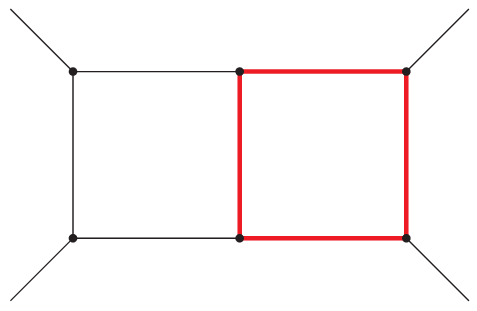}
\end{center}
\caption{
Two examples of planar double-box Feynman integrals with internal masses.
Black lines denote massless particles, red lines denote particles with a mass $m$.
}
\label{fig_massive_double_box}
\end{figure}
Massive propagators are indicated by red lines in the left diagram of figure~\ref{fig_massive_double_box}.
The system consists of $29$ master integrals.
This integral was first analytically computed in ref.~\cite{Caron-Huot:2014lda} via an $\eps$-factorised differential equation.

Thirdly, we studied a further planar double-box integral with one internal mass.
The massive propagators are indicated by red lines in the right diagram of figure~\ref{fig_massive_double_box}.
This family differs from the previous one by the mass assignments for the internal propagators, and consists of $32$ master integrals.
The first analytic computation was presented in ref.~\cite{Becchetti:2017abb}, again via an $\eps$-factorised differential equation.

These examples have been chosen as they provide sectors with up to four master integrals and several square roots.
In all cases we were able to realise self-duality and Galois symmetries.

\section{Beyond dlog-forms}
\label{sect:beyond_dlog}

\begin{figure}
\begin{center}
\includegraphics[scale=0.8]{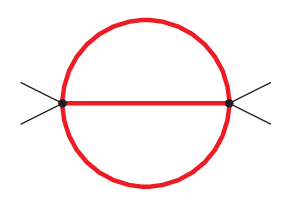}
\hspace*{10mm}
\includegraphics[scale=0.8]{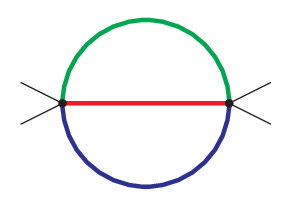}
\end{center}
\caption{
The Feynman diagram of the equal-mass sunrise integral (left) and the unequal-mass sunrise integral (right).
Coloured lines denote massive particles and different colours correspond to different masses.
}
\label{fig_sunrise}
\end{figure}

It is well-known that there exist Feynman integrals whose differential equations involve differential one-forms beyond
dlog-forms with algebraic arguments.
A typical example are elliptic Feynman integrals, and more generally, Calabi--Yau Feynman integrals.
In this section we discuss Galois symmetries for elliptic Feynman integrals.
In the elliptic case there is an additional integer number, called the modular weight $k$.
The known $\eps$-factorised differential equations in the elliptic case have the property
that each entry of the differential equation matrix $A$ has a unique modular weight $k$.
A Galois symmetry permutes roots of a polynomial equation, it does not change the modular weight.
Hence, Galois symmetries cannot relate entries of different modular weight.
For this reason, there is no Galois symmetry in the simplest elliptic Feynman integral, 
the equal-mass two-loop sunrise integral, shown in fig~\ref{fig_sunrise}.
However, there can be Galois symmetries relating entries of the same modular weight.
As an example of a Galois symmetry in the elliptic case we discuss the unequal-mass sunrise integral, also shown in fig.~\ref{fig_sunrise} and a two-loop non-planar three-point function shown in fig.~\ref{fig_triangle_a}.

\subsection{The equal-mass sunrise integral}
\label{sect:equal_mass_sunrise}

The family of the equal-mass two-loop sunrise integral consists of two sectors: one sector with one master integral corresponding to
the product of two one-loop tadpole integrals and a second sector with two master integrals corresponding to the sunrise topology.
The differential equation can be brought into an $\eps$-factorised form and reads \cite{Adams:2018yfj}
\bq
\label{dgl_sunrise}
 d I & = & \eps A I,
 \;\;\;\;\;\;
 A
 \; = \;
 \left( \begin{array}{ccc}
 0 & 0 & 0 \\
 0 & \cellcolor{yellow} \omega_2 & \omega_0 \\
 \omega_3 & \omega_4 & \cellcolor{yellow} \omega_2 \\
 \end{array} \right).
\eq
The sunrise integral depends on one kinematic variable, which we may take to be the modular parameter $\tau$.
In this variable we have
\bq
 \omega_k
 & = &
 f_k(\tau) \; \left(2\pi i\right) d\tau,
\eq
where $f_k(\tau)$ is a modular form of modular weight $k$.
The matrix $A$ in eq.~(\ref{dgl_sunrise}) clearly has the self-duality symmetry, which is the statement that
\bq
 a_{22} \; = \; a_{33} \; = \; \omega_2.
\eq
However, the modular form $f_0(\tau)$ appearing in $a_{23}=\omega_0$ is of modular weight $0$,
while the modular form $f_4(\tau)$ appearing in $a_{32}=\omega_4$ is of modular weight $4$.
There cannot be a Galois symmetry relating $f_0(\tau)$ to $f_4(\tau)$.

\subsection{The unequal-mass sunrise integral}

The situation is different for the unequal-mass sunrise integral.
This is a system with seven master integrals.
There are three sectors with one master integral each, corresponding to products of two one-loop tadpole
integrals.
In addition, there exists a sector with four master integrals corresponding to the sunrise topology.
In the latter sector we may choose the master integrals such that two of them reduce to the master integrals
of the equal-mass case in the equal-mass limit. As in the equal-mass case there is no Galois symmetry
relating these two.
The other two master integrals can be chosen such that they vanish in the equal-mass limit.
More importantly, the entries in the differential equation related to these master integrals have the same modular weights.
A redefinition of these will realise a Galois symmetry.
In the following we will follow the notation of ref.~\cite{Bogner:2019lfa}.
We have an $\eps$-factorised differential equation
\bq
 d J
 & = & 
 \eps A J,
\eq
with a $(7 \times 7)$-matrix $A$. The master integrals $J$ are defined in eq.~(78) of ref.~\cite{Bogner:2019lfa}.
The modular weights of the non-zero entries of $A$ are
\bq
 \left( \begin{array}{ccccccc}
 2 & - & - & - & - & - & - \\
 - & 2 & - & - & - & - & - \\
 - & - & 2 & - & - & - & - \\
 - & - & - & 2 & 1 & 1 & 0 \\
 2 & 2 & 2 & 3 & 2 & 2 & 1 \\
 2 & 2 & 2 & 3 & 2 & 2 & 1 \\
 3 & 3 & 3 & 4 & 3 & 3 & 2 \\
 \end{array} \right)
\eq
and we see that corresponding entries related to $J_5$ and $J_6$ have the same modular weights.
Changing to a new basis
\bq
 J_5' \; = \; \frac{1}{12} \sqrt{3} J_5 + \frac{i}{4} J_6,
 & &
 J_6' \; = \; \frac{1}{12} \sqrt{3} J_5 - \frac{i}{4} J_6,
\eq
and $J_k'=J_k$ for all other $k$ will realise self-duality and the Galois symmetry
\bq
 J_6' & = & \sigma_{-1}\left(J_5'\right).
\eq
We have again introduced two square roots
\bq
 s_3 \; = \; \sqrt{3},
 & &
 s_{-1} \; = \; i.
\eq
In the new basis $J'$ we have again an $\eps$-factorised differential equation
\bq
 d J' & = & \eps A' J',
 \;\;\;\;\;\;
 A'
 \; = \;
 \left( \begin{array}{ccccccc}
 a_{11}' & 0 & 0 & 0 & 0 & 0 & 0 \\
 0 & a_{22}' & 0 & 0 & 0 & 0 & 0 \\
 0 & 0 & a_{33}' & 0 & 0 & 0 & 0 \\
 0 & 0 & 0 & \cellcolor{red} a_{44}' & \cellcolor{yellow} a_{45}' & \cellcolor{yellow} a_{46}' & a_{47}' \\
 \cellcolor{magenta} a_{51}' & \cellcolor{purple} a_{52}' & \cellcolor{pink} a_{53}' & \cellcolor{lime} a_{54}' & \cellcolor{cyan} a_{55}' & \cellcolor{orange} a_{56}' & \cellcolor{yellow} a_{57}' \\
 \cellcolor{magenta} a_{61}' & \cellcolor{purple} a_{62}' & \cellcolor{pink} a_{63}' & \cellcolor{lime} a_{64}' & \cellcolor{orange} a_{65}' & \cellcolor{cyan} a_{66}' & \cellcolor{yellow} a_{67}' \\
 a_{71}' & a_{72}' & a_{73}' & a_{74}' & \cellcolor{lime} a_{75}' & \cellcolor{lime} a_{76}' & \cellcolor{red} a_{77}' \\
 \end{array} \right).
\eq
The matrix $A'$ satisfies now the self-duality relations
\bq
 a_{77}' \; = \; a_{44}',
 \;\;\;\;\;\;
 a_{66}' \; = \; a_{55}',
 \;\;\;\;\;\;
 a_{67}' \; = \; a_{45}',
 \;\;\;\;\;\;
 a_{57}' \; = \; a_{46}',
 \;\;\;\;\;\;
 a_{76}' \; = \; a_{54}',
 \;\;\;\;\;\;
 a_{75}' \; = \; a_{64}',
\eq
and in addition the Galois symmetries
\bq
 a_{65}' \; = \; \sigma_{-1}\left(a_{56}'\right),
 \;\;\;\;\;\;
 a_{46}' \; = \; \sigma_{-1}\left(a_{45}'\right),
 \;\;\;\;\;\;
 a_{64}' \; = \; \sigma_{-1}\left(a_{54}'\right),
\eq
and
\bq
 a_{6j}' \; = \; \sigma_{-1}\left(a_{5j}'\right)
 & & 
 \mbox{for} \; j \in \{1,2,3\}.
\eq

\subsection{A non-planar elliptic three-point function}
\label{sect:triangle_a}

\begin{figure}
\begin{center}
\includegraphics[scale=1.0]{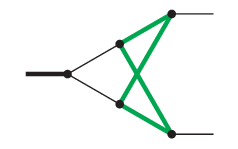}
\end{center}
\caption{
An example of a two-loop non-planar elliptic three-point function.
Thin black lines denote massless particles, 
green lines denote particles with a
mass $m$.
}
\label{fig_triangle_a}
\end{figure}

We consider the two-loop non-planar three-point function shown in fig.~\ref{fig_triangle_a}.
This family consists of $11$ master integrals, grouped into $8$ sectors.
There are three sectors with two master integrals each. 
Among the sectors with two master integrals there is the top sector, which is elliptic.
An $\eps$-factorised differential equation has been given in ref.~\cite{Jiang:2023jmk},
and we follow to a large extent the notation of this paper.
As we always order the sectors from the simplest to the most complicated sector, we start from
a basis $I$ with
\bq
 I_3 \; = \; M_{8},
 \;\;\; 
 I_4 \; = \; M_{9},
 \;\;\; 
 I_6 \; = \; M_{5},
 \;\;\; 
 I_7 \; = \; M_{6},
 \;\;\; 
 I_{10} \; = \; M_{1},
 \;\;\; 
 I_{11} \; = \; M_{2},
\eq
and $I_k=M_{12-k}$ for all others.
The master integrals $M_k$ are defined in ref.~\cite{Jiang:2023jmk}.
This example has two kinematic square roots
\bq
 r_{1} \; = \; \sqrt{1+4y},
 & &
 r_{-1} \; = \; \sqrt{1-4y},
\eq
where $y$ is the kinematic variable defined in ref.~\cite{Jiang:2023jmk}.
In order to realise self-duality and Galois symmetries,
we introduce
\bq
 s_3 \; = \; \sqrt{3},
 & &
 s_6 \; = \; \sqrt{6}
\eq
and set
\bq
 J_3 \; = \; I_3 + \frac{\sqrt{6}}{3} I_4,
 & &
 J_4 \; = \; I_3 - \frac{\sqrt{6}}{3} I_4,
 \nonumber \\
 J_6 \; = \; I_7 + \frac{\sqrt{3}}{3} I_6,
 & &
 J_7 \; = \;  I_7 - \frac{\sqrt{3}}{3} I_6.
\eq
We have the Galois symmetries
\bq
 J_4 \; = \; \rho_{1}\left(J_3\right) \; = \; \sigma_6\left(J_3\right),
 & &
 J_7 \; = \; \rho_{-1}\left(J_6\right) \; = \; \sigma_3\left(J_6\right).
\eq
There is no Galois symmetry for the elliptic sector consisting of the master integrals $J_{10}$ and $J_{11}$
(for the same reasons as discussed in section~\ref{sect:equal_mass_sunrise}).
In the basis $J$ we have again an $\eps$-factorised differential equation
\bq
 d J & = & \eps A J,
\eq
where the $(11 \times 11)$-matrix $A$ has the structure (we use again hexadecimal indices)
\bq
 A & = &
 \left( \begin{array}{ccccccccccc}
 0 & 0 & 0 & 0 & 0 & 0 & 0 & 0 & 0 & 0 & 0 \\
 0 & a_{22} & 0 & 0 & 0 & 0 & 0 & 0 & 0 & 0 & 0 \\
 \cellcolor{pink} a_{31} & 0 & \cellcolor{red} a_{33} & \cellcolor{orange} a_{34} & 0 & 0 & 0 & 0 & 0 & 0 & 0 \\
 \cellcolor{pink} a_{41} & 0 & \cellcolor{orange} a_{43} & \cellcolor{red} a_{44} & 0 & 0 & 0 & 0 & 0 & 0 & 0 \\
 0 & 0 & \cellcolor{magenta} a_{53} & \cellcolor{magenta} a_{54} & 0 & 0 & 0 & 0 & 0 & 0 & 0 \\
 \cellcolor{Goldenrod} a_{61} & \cellcolor{yellow} a_{62} & 0 & 0 & 0 & \cellcolor{green} a_{66} & \cellcolor{lime} a_{67} & 0 & 0 & 0 & 0 \\
 \cellcolor{Goldenrod} a_{71} & \cellcolor{yellow} a_{72} & 0 & 0 & 0 & \cellcolor{lime} a_{76} & \cellcolor{green} a_{77} & 0 & 0 & 0 & 0 \\
 0 & 0 & 0 & 0 & a_{85} & \cellcolor{ForestGreen} a_{86} & \cellcolor{ForestGreen} a_{87} & a_{88} & 0 & 0 & 0 \\
 0 & 0 & 0 & 0 & a_{95} & 0 & 0 & 0 & 0 & 0 & 0 \\
 0 & 0 & 0 & 0 & 0 & 0 & 0 & 0 & 0 & \cellcolor{blue} a_{AA} & a_{AB} \\
 0 & 0 & \cellcolor{violet} a_{B3} & \cellcolor{violet} a_{B4} & a_{B5} & \cellcolor{Aquamarine} a_{B6} & \cellcolor{Aquamarine} a_{B7} & a_{B8} & 0 & a_{BA} & \cellcolor{blue} a_{BB} \\
 \end{array} \right).
\eq
Entries with the same background colour are related by a symmetry.
Note that there is no Galois symmetry relating $a_{AB}$ to $a_{BA}$. 
The former entry has modular weight $0$, the latter modular weight $4$.

\section{Conclusions}
\label{sect:conclusions}

In this paper we studied the structure of the differential equation for families of Feynman integrals.
We showed that often the connection matrix $A$ has additional symmetries, which can be realised
by a redefinition of the master integrals through a constant $\mathrm{GL}(\NF,{\mathbb C})$-transfor\-ma\-tion.
The symmetries we studied were self-duality and Galois symmetry.
Self-duality is the statement that blocks on the diagonal are symmetric with respect to the anti-diagonal.
In all examples we presented we were able to find a transformation which achieves self-duality.

In addition to self-duality there can be Galois symmetries present.
Galois symmetry relates two master integrals 
through the action of a Galois group: $I_2=\sigma(I_1)$.
Galois symmetries can be expected in sectors where the definition of the master integrals for an
$\eps$-factorised differential equation involves square roots.
Surprisingly, also sectors not related to any square root with dependence on
the kinematic variables may exhibit Galois symmetries.
The requirement of self-duality can introduce constant square roots like $\sqrt{3}$ and the Galois symmetry
in the latter case is the conjugation $\sqrt{3} \rightarrow - \sqrt{3}$.
Galois symmetries may or may not exist on top of self-duality.
We presented many examples with Galois symmetries, but we also pointed out two examples, where there is
no Galois symmetry on top of self-duality: These were the examples of the two-loop planar massless double-box
integral and the two-loop equal-mass sunrise integral.
In the former case we showed that we still have a limit Galois symmetry on top of self-duality.
In the latter case we do not expect Galois symmetries, as Galois symmetries cannot 
relate quantities of different modular weight.
To the other extreme, we also find examples in which a Galois symmetry
can be realised in a number of different ways. Moreover, there exist cases with symmetries
involving arbitrary roots, including perfect squares. Thus the presence of Galois symmetry is not
necessarily linked with a specific or even any field extension 
(see also the discussion in section~\ref{sect:rationalisation}).

Clearly, self-duality and Galois symmetries give only non-trivial relations in sectors with
two or more master integrals
and therefore only affect more complicated Feynman integrals.
This might be an explanation why these symmetries have not been noticed up to now.
In this paper we presented strong evidence for the ubiquity of self-duality and Galois symmetries in Feynman integrals.
It would be interesting to understand the exact conditions under which these symmetries can be realised,
and to which extent they are unique.
This is left for future work. 

\subsection*{Note added}

While this paper was under review, ref.~\cite{Duhr:2024xsy} appeared on the arXiv, 
providing an explanation of the self-duality symmetry we observed in terms of twisted
cohomology on the maximal cut.

\subsection*{Acknowledgements}

K.W. is very grateful for the hospitality provided by the Institut f\"ur Physik, Mainz.
K.W. is supported by the Helmholtz-OCPC International Postdoctoral Exchange Fellowship Program.
This work has been supported by the 
Cluster of Excellence Precision Physics, Fundamental Interactions, and Structure of
Matter (Grant No.~EXC - 2118 - 390831469) 
and by the Cluster of Excellence ORIGINS (Grant No.~EXC - 2094 - 390783311),
both funded by the German Research Foundation (DFG) within
the German Excellence Strategy.
The work of X.X. is supported by the European Research Council (ERC) under the European Union’s Horizon
2022 Research and Innovation Program (ERC Advanced Grant agreement No.101097780, EFT4jets). Views and opinions
expressed are however those of the authors only and do not necessarily reflect those of the European Union or
the European Research Council Executive Agency. Neither the European Union nor the granting authority can be
held responsible for them.

{\footnotesize
\bibliography{/home/stefanw/notes/biblio}
\bibliographystyle{/home/stefanw/latex-style/h-physrev5}
}

\end{document}